\documentclass[
  twocolumn,
  prl,
  showpacs,
  amsmath,
  amssymb,
  superscriptaddress,
  floatfix
]{revtex4}

\usepackage{bm}
\usepackage{graphicx}
\usepackage{color}
\usepackage{hyperref}

\usepackage{ulem}

\DeclareMathOperator{\tr}{tr}
\DeclareMathOperator{\sgn}{sgn} \DeclareMathOperator{\Imag}{Im}
\DeclareMathOperator{\Real}{Re} 
 
\DeclareMathOperator{\lit}{li_4}
\DeclareMathOperator{\liq}{li_2}

\newcommand{\be}{\begin{equation}}
\newcommand{\ee}{\end{equation}}
\newcommand{\beml}{\begin{subequations}}
\newcommand{\eml}{\end{subequations}}
\newcommand{\bea}{\begin{eqnarray}}
\newcommand{\eea}{\end{eqnarray}}
\newcommand{\ba}{\begin{array}}
\newcommand{\ea}{\end{array}}
\newcommand{\bpm}{\begin{pmatrix}}
\newcommand{\epm}{\end{pmatrix}}

\DeclareMathOperator{\im}{Im}
\DeclareMathOperator{\re}{Re}

\begin{document}

\title{Multifractality at Anderson transitions with Coulomb interaction}

\author{I.S.~Burmistrov}

\affiliation{ L.D. Landau Institute for Theoretical Physics, Kosygina
  street 2, 119334 Moscow, Russia}

\author{I.V.~Gornyi}
\affiliation{
 Institut f\"ur Nanotechnologie, Karlsruhe Institute of Technology,
 76021 Karlsruhe, Germany
}
\affiliation{
 A.F.~Ioffe Physico-Technical Institute,
 194021 St.~Petersburg, Russia.
}

\author{A.D.~Mirlin}
\affiliation{
 Institut f\"ur Nanotechnologie, Karlsruhe Institute of Technology,
 76021 Karlsruhe, Germany
}
\affiliation{
 DFG Center for Functional Nanostructures,
 Karlsruhe Institute of Technology, 76128 Karlsruhe, Germany
}
\affiliation{
 Institut f\"ur Theorie der kondensierten Materie,
 Karlsruhe Institute of Technology, 76128 Karlsruhe, Germany
}
\affiliation{
 Petersburg Nuclear Physics Institute,
 188300 St.~Petersburg, Russia.
}

\begin{abstract}
We explore mesoscopic fluctuations and correlations of
the local density of states (LDOS)
near localization transition in a disordered interacting electronic system. It
is shown that the LDOS multifractality survives in the presence of Coulomb
interaction. We calculate the spectrum of multifractal dimensions in
$2+\epsilon$ spatial dimensions and show that it differs from that in the
absence of
interaction. The multifractal character of  fluctuations and correlations
of the LDOS can be studied
experimentally by scanning tunneling microscopy of two-dimensional and
three-dimensional disordered structures.

\end{abstract}

\pacs{
72.15.Rn , \,
71.30.+h , \,
73.43.Nq 	\,
}

\maketitle

Fifty five years after its discovery \cite{Anderson58}, Anderson localization
remains an actively developing field \cite{AL50}. One of central directions of
the current research is the physics of Anderson localization transitions
\cite{Evers08}. These include both metal-insulator transitions and quantum Hall
plateau transitions (and, more generally, transitions between different phases
of topological insulators). Such transitions have been experimentally observed
and studied in a variety of semiconductor structures \cite{semicond}. Recent
discoveries of graphene \cite{graphene} and
time-reversal-invariant topological insulator materials \cite{topins}
have further broadened the arena for their experimental exploration. In addition
to electronic
systems, there is a number of further experimental realizations of Anderson
localization, including localization of light \cite{wiersma97}, cold atoms
\cite{BEC-localization}, ultrasound \cite{faez09}, and optically driven atomic
systems \cite{lemarie10}.

Anderson transitions are quantum phase transitions and are characterized by
critical scaling of various physical observables. A particularly remarkable
property of Anderson transitions is the
multifractality of critical wave functions which implies their very strong
fluctuations. Specifically, at the critical point the wave-function moments
or equivalently, the
averaged participation ratios
$\langle P_q\rangle=\langle\int d^dr|\psi({\bf r})|^{2q}\rangle$
show anomalous multifractal scaling with respect to the
system size $L$,
\begin{equation}\label{e1}
 L^d \langle |\psi({\bf r})|^{2q} \rangle \propto L^{-\tau_q}, \qquad \tau_q =
d(q-1) +  \Delta_q,
\end{equation}
where $d$ is the spatial dimension, $\langle\ldots\rangle$ denotes the
averaging over disorder, and $\Delta_q$ are anomalous multifractal
exponents distinguishing the critical point from a conventional metallic
phase, where $\Delta_q\equiv 0$. We refer the reader to
Ref.~\cite{Evers08} for an overview of this research area. Very recently, a
complete classification of observables characterizing critical wave functions
(that includes multifractal moments (\ref{e1}) as a ``tip of the iceberg'')
was developed \cite{gruzberg13}.

The above results on multifractality have been obtained for
{\it non-interacting} disordered systems. In the case of broken spin invariance
they remain valid in
the presence of short-range (e.g., screened by external gate) electron-electron
interaction which, in this case, is irrelevant in the renormalization-group
(RG) sense \cite{note2}. On the
other hand, the long-range ($1/r$) Coulomb interaction
is RG relevant and may have strong impact on localization properties of the
system (see Refs. \cite{finkelstein90,belitz94} for review). In particular,
it induces a
metal-insulator transition in (otherwise localized) two-dimensional systems with
preserved spin and time-reversal invariances \cite{punnoose05}.
Further, the Coulomb interaction induces a strong suppression of the local
density of states (LDOS) $\rho(E)$ near zero energy $E$
(counted from the chemical potential) \cite{AA1979AAL1980,ES75SE84}. The
LDOS can be measured in a tunneling experiment, and this phenomenon is known as
zero-bias anomaly (ZBA). Specifically, in a two-dimensional (2D) weakly disordered
system the disorder-averaged LDOS behaves as
\cite{Fin1983,Castellani1984,finkelstein90,belitz94,nazarov89,levitov97,kamenev99}
\begin{equation}
\langle \rho(E) \rangle \propto \exp\left\{-{1\over 4\pi g} \log^2 |E|
\right\} ,
\label{e2}
\end{equation}
where $g$ is the dimensionless (measured in units $e^2/h$) conductivity.
The physics of a 2D disordered systems is closely related to the behavior at
Anderson transition, since $d=2$ is a logarithmic
(lower critical) dimension. The unconventional behavior (\ref{e2}) with squared
logarithm in the exponential (rather than with a simple logarithm that would
yield a power law, as normally expected for critical behavior) is related to
the fact that the LDOS is affected by gauge-type phase fluctuations that yield
a suppression of Debye-Waller type. For the Anderson transition in
$d=2+\epsilon$ dimensions (with $\epsilon \ll 1$ allowing a parametric control
of the theory) in systems with broken time-reversal and/or spin symmetries, one
of the logarithmic factors in Eq.~(\ref{e2}) transforms into a factor $\sim
1/\epsilon$ \cite{finkelstein90,belitz94}. Since the critical conductance
$g_*$ is of order $1/\epsilon$ as well, this yields
\begin{equation}
\langle \rho(E) \rangle \propto |E|^\beta, \qquad \beta = O(1),
\label{e3}
\end{equation}
with precise value of the critical exponent $\beta$ depending on the
symmetry class. Specifically,
up to corrections of order $\epsilon$, one finds $\beta\simeq 1/2$,
$1/[4(1-\ln 2)]$, and 1, for the problems with magnetic impurities,
magnetic field, and
spin-orbit scattering, respectively \cite{finkelstein90,belitz94}.
In view of a combination of disorder and interaction physics,
such metal-insulator transitions are often called Mott-Anderson (or
Anderson-Mott) transitions. We remind
that in the absence of interaction the LDOS is uncritical, $\beta=0$, in
conventional (Wigner-Dyson) symmetry classes.

We are thus facing the following important question: Does multifractality
survive in the presence of Coulomb interaction between electrons?
The goal of this paper is to answer this question. Specifically, we will show
that on top of the ZBA suppression of the average $\langle\rho(E)\rangle$, the
LDOS of a strongly interacting critical system does show multifractal
fluctuations and correlations. We will also calculate the corresponding spectrum
of anomalous dimensions in $2+\epsilon$ spatial dimensions up to the
two-loop order and demonstrate that it differs from that of a
non-interacting system.

Note that the question addressed in this Letter is of direct
experimental relevance. In
particular, a recent work \cite{richardella10} performed scanning tunneling
microscopy (STM) of a magnetic semiconductor Ga$_{1-x}$Mn$_x$As near
metal-insulator transition and detected LDOS fluctuations and correlations of
mutlifractal character. Strong fluctuations of the LDOS in a strongly
disordered
3D system (presumably, on the insulating side of the transition) have been also
observed in Ref.~\cite{morgenstern02}. Further, recent
works on STM of 2D semiconductor systems and graphene
\cite{morgenstern-2D} demonstrated the feasibility to explore
fluctuations and correlations of the LDOS also near the quantum Hall
transitions.
Finally, strong spatial fluctuations of the LDOS have been also
detected near the superconductor-insulator transition in
disordered films \cite{sacepe08} that is known to have much in common
with the metal-insulator transition.

We turn now to the presentation of our results. The LDOS is formally defined as
an imaginary part of the single-particle Green function,
$\rho(E,\bm{r})=(-1/\pi)\im G(E;\bm{r},\bm{r})$. We find that near the
Anderson localization transition
the moments of LDOS normalized to its average show multifractal scaling,
\begin{equation}
\Bigl \langle [\rho(E,\bm{r})]^q \Bigr  \rangle / \langle \rho(E)  \rangle^q
\sim (\mathcal{L}/l )^{-\Delta_q},
\label{e4}
\end{equation}
where $l$ denotes a microscopic length scale of the order of elastic
scattering mean free path and $\mathcal{L}=\min\{\xi,L_\phi,L\}$ is the
shorter of the three lengths:
the localization (correlation) length $\xi$, the dephasing length $L_\phi$, and
the system size $L$ \cite{note1}. The correlation length diverges at the
transition point in a power-law fashion, $\xi\sim |g-g_*|^{-\nu}$, with an
exponent $\nu$. Further, the dephasing length (controlled by inelastic
scattering processes) diverges
at zero energy (we remind that all energies are counted from the chemical
potential), $L_\phi\sim |E|^{-1/z}$, with a dynamical exponent $z$.

The power-law scaling (\ref{e4}) of the normalized LDOS moments is governed by
a set
of exponents $\Delta_q$. These exponents control also spatial
power-law correlations of the LDOS at scales $R<\mathcal{L}$. (At large
distances, $R\gg\mathcal{L}$, the LDOS become essentially uncorrelated.) In
particular, the correlation function of two LDOS at different points shows at
$l<R<\mathcal{L}$ the following scaling:
\begin{equation}
 \label{e5}
\Bigl \langle   \rho(E, \bm{r})  \rho(E, \bm{r}+\bm{R}) \Bigr \rangle  \Bigl /
\langle \rho(E) \rangle^2 \sim
\left (\mathcal{L}\over R\right)^{\eta},
\end{equation}
where $\eta=-\Delta_2$.
Correlations between the LDOS at different energies have analogous
scaling properties,
\begin{equation}
\frac{\bigl \langle   \rho(E, \bm{r})  \rho(E+\omega, \bm{r}+\bm{R}) \bigr
\rangle}{\langle \rho(E) \rangle \langle \rho(E+\omega)} \sim \left
(\frac{L_\omega}{R}\right)^{\eta}, \quad R <
L_\omega ,
\label{e6}
\end{equation}
where $L_\omega\sim\omega^{-1/z}$ and it is assumed that
$L_\omega<\mathcal{L}$.

To derive the above results, we use the
non-linear $\sigma$ model (NL$\sigma$M) field-theoretical approach to
interacting disordered systems \cite{finkelstein90,belitz94}. To keep the
analysis parametrically under control, we consider the Anderson transition in
$d=2+\epsilon$ dimensions, where the critical conductance
$g_*$ is large (i.e. the corresponding $\sigma$ model coupling $t_*=1/\pi
g_*$ is weak). This allows us to obtain an $\epsilon$ expansion for critical
exponents. In spirit of usual ideology of critical phenomena, it is expected
that the scaling results  \eqref{e4}-\eqref{e6} are of general
validity and hold also at strong-coupling critical points of
Coulomb-interacting disordered systems, such as metal-insulator transitions in
3D or quantum Hall plateau transitions.

We consider a system of disordered fermions with Coulomb
interaction in the absence of time reversal and spin rotational symmetries,
which corresponds to the symmetry class ``MI(LR)'' in terminology of
Ref.~\cite{belitz94}). The RG analysis of the Anderson metal-insulator
transition in $d=2+\epsilon$ dimensions for this symmetry class was
developed up to two-loop order in Refs~\cite{baranov99, baranov02}.
Renormalization of the dimensionless
conductance $g$ is
governed by the following $\beta$-function \cite{baranov02}
\begin{equation}
-\frac{dt}{d\ln y} = \beta(t) = \epsilon t - 2 t^2 - 4 A t^3 + O(t^4) ,
\label{eq2E1}
\end{equation}
where $y$ is the running RG length scale, $t=1/\pi g$, and $A\approx 1.64$.
The condition $\beta(t_*)=0$ determines the position of the critical point:
$t_*=(\epsilon/2)(1-A\epsilon)+O(\epsilon^3)$ (and thus the critical
conductance $g_*=1/\pi t_*$). Further, the localization
length exponent is determined by the derivative of the $\beta$-function
at the fixed point, $\nu=-1/\beta^\prime(t_*)=1/\epsilon-A+O(\epsilon)$.
The dynamical exponent connecting the energy and length scaling
at criticality is also known up to the two-loop order:
$z=2+\epsilon/2+(2A-\pi^2/6-3)\epsilon^2/4+O(\epsilon^3)$ \cite{baranov99}.

To determine the scaling of LDOS moments, we translate the corresponding
correlation functions into the NL$\sigma$M language. We use the Matsubara
version of the interacting $\sigma$ model, and introduce replicas in order to
perform the disorder averaging.
A detailed two-loop RG analysis (see Supporting Material \cite{EPAPS})
demonstrates that the scaling behavior of moments of the normalized LDOS,
$[\rho(E,\bm{r})/\langle\rho(E)\rangle]^q$, is governed by the anomalous
dimensions
\begin{equation}
\zeta_q(t) = \frac{q(1-q) t}{2}\left [ 1 +  \left (2-\frac{\pi^2}{6}\right ) t
\right ] +O(t^3) .
\label{eq2E2}
\end{equation}
This proves the anomalous scaling (\ref{e4}) and
determines the multifractal exponents at the critical point:
\begin{equation}
\Delta_q = \zeta_q(t_*) =
\frac{q(1-q)\epsilon}{4}\Bigl [1 + \left (1-A-\frac{\pi^2}{12}\right ) \epsilon \Bigr ]+O(\epsilon^3) .
\label{eq2E3}
\end{equation}
An extension of this analysis onto correlation functions of LDOS at different
spatial points and/or energies yields Eqs.~(\ref{e5}), (\ref{e6}) and their
generalizations onto higher correlation functions \cite{EPAPS}.

\begin{figure}[t]
\centerline{a) \includegraphics[height=15mm]{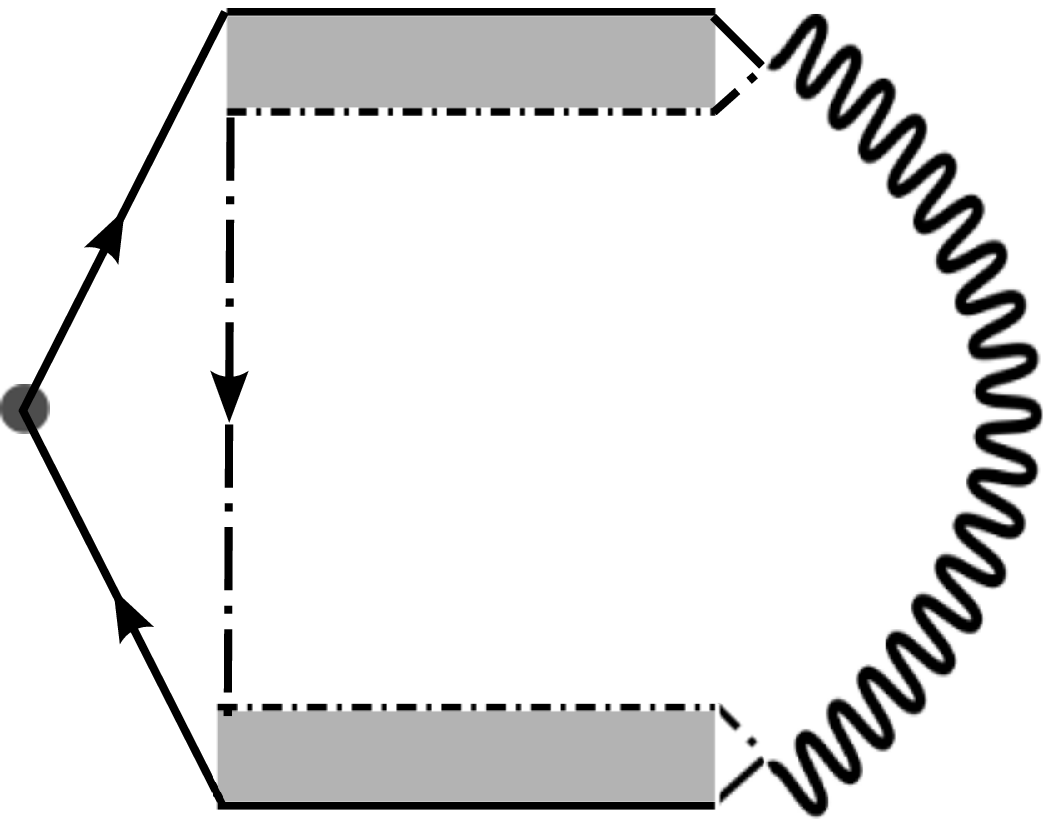}\hspace{0.1cm} b) \includegraphics[height=15mm]{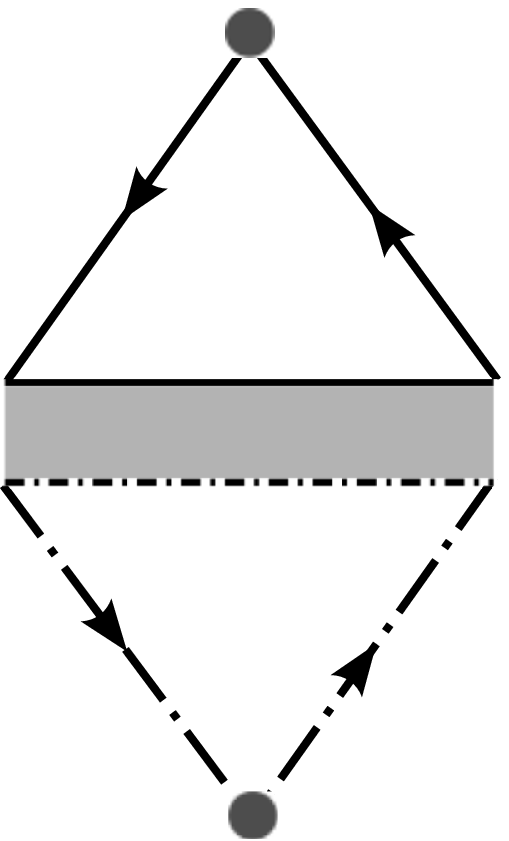}\hspace{0.1cm} c) \includegraphics[height=15mm]{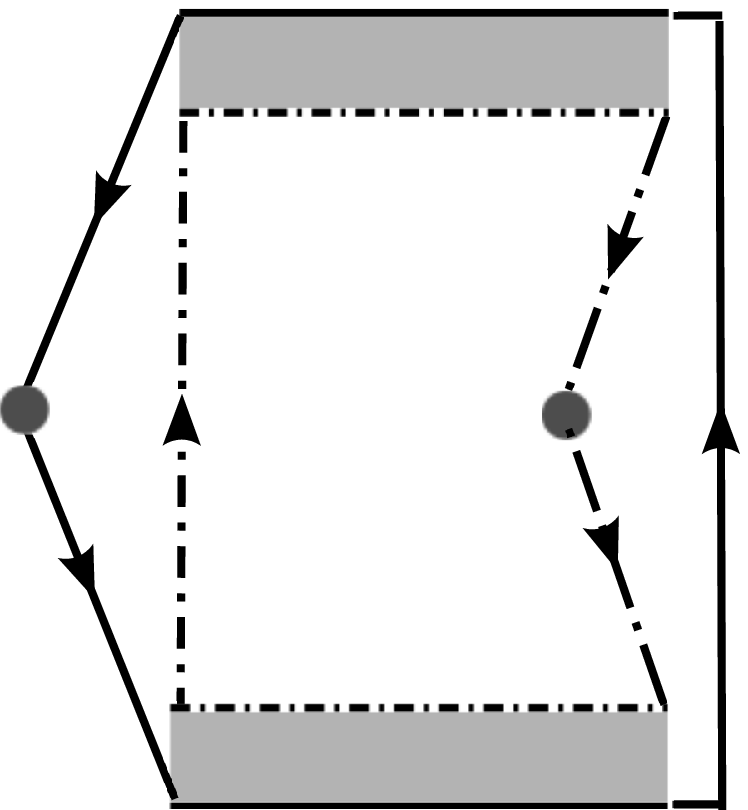}\hspace{0.1cm} d) \includegraphics[height=15mm]{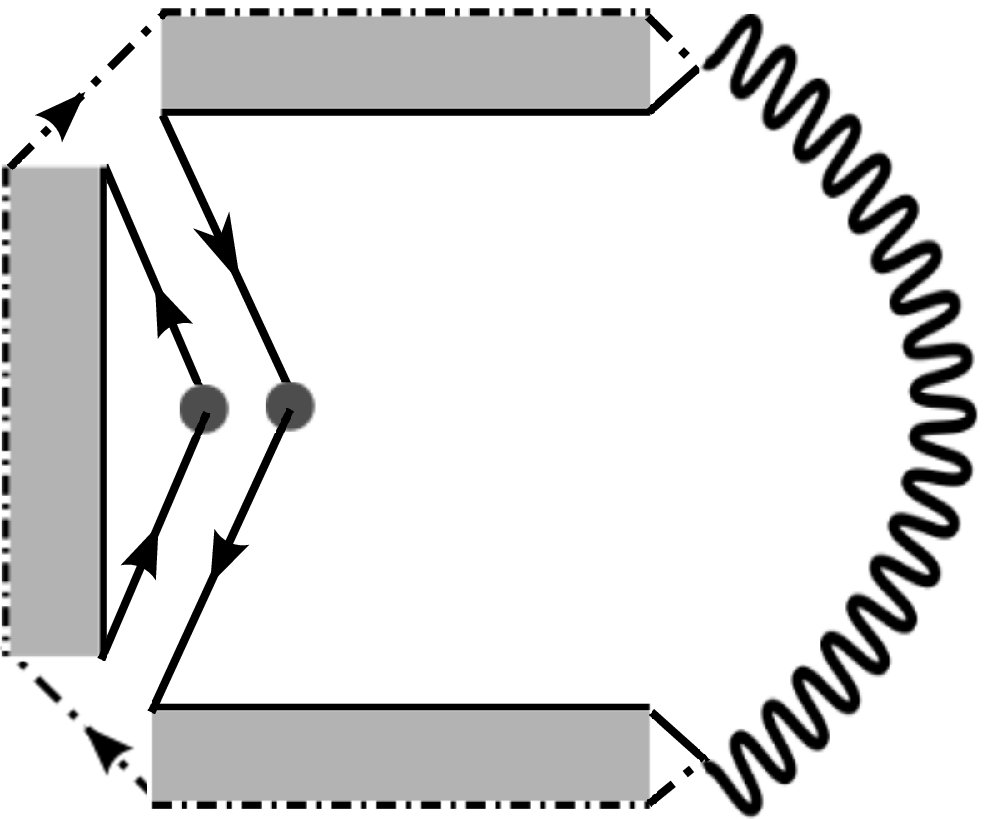}
}
\vspace{0.6cm}
\centerline{e) \includegraphics[height=3.9mm]{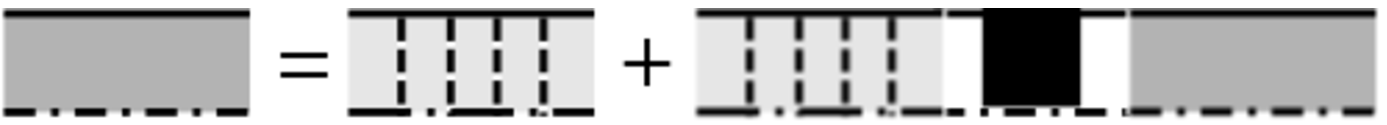}\hspace{0.1cm}
f) \includegraphics[height=3.9mm]{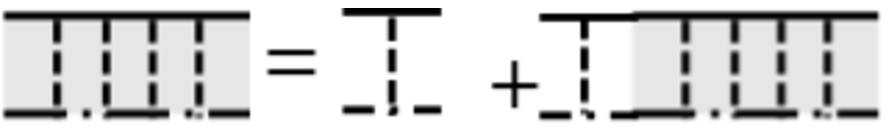}
}
\vspace{0.3cm}
\centerline{g) \includegraphics[width=75mm]{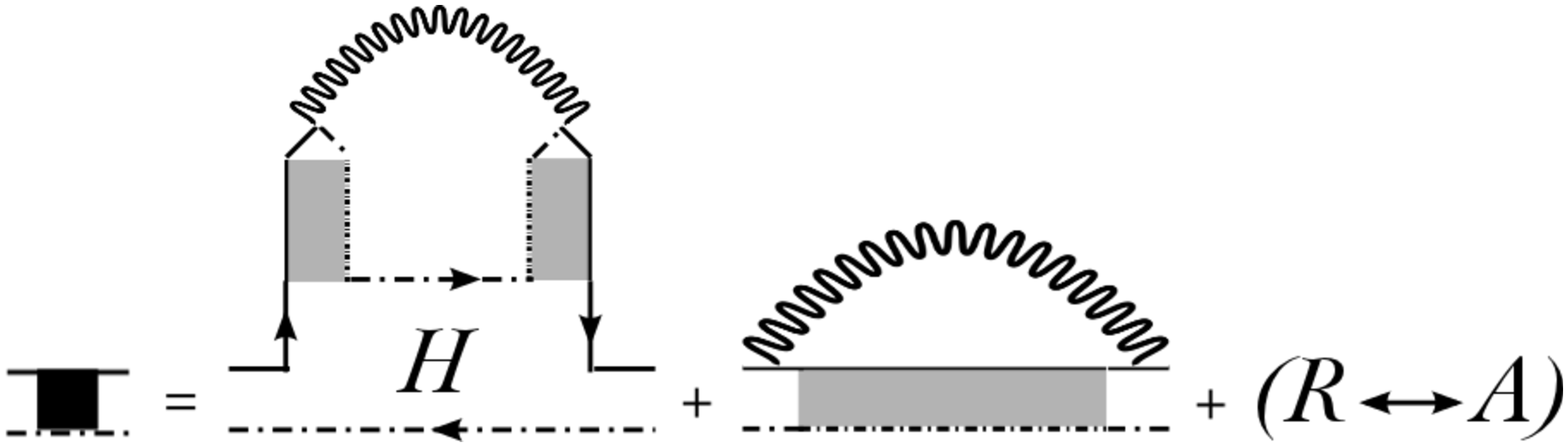}
}
\vspace{0.3cm}
\centerline{h) \includegraphics[height=10mm]{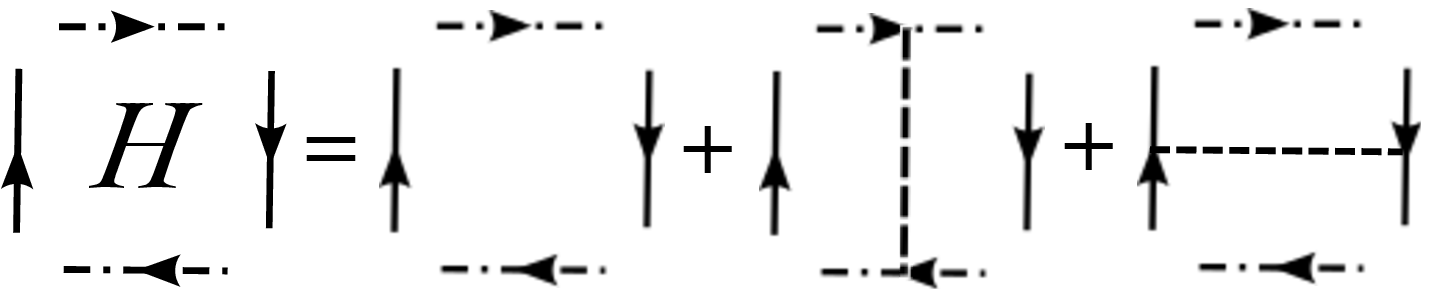}
}
\caption{Representative diagrams for the average LDOS (a) and to the correlation function of two LDOS $\langle \rho(E, \bm{r})  \rho(E+\omega, \bm{r}+\bm{R}) \rangle$ (b-d). The retarded (advanced) single-particle Green function is denoted by solid (dashed-dotted) line. Wavy solid line denotes the dynamically screened Coulomb interaction. Shaded rectangular is a representation for the diffuson (e) with self-energy due to interaction shown in (g). The shaded rectangular with dashed lines stands for a bare diffuson (f). The impurity line is denoted by dashed line. White rectangular with symbol {\it `H'} stands for the Hikami box shown in (h). }
\label{Figure1}
\end{figure}

To illustrate the origin of obtained fluctuations and correlations of the LDOS,
we show in Fig.~\ref{Figure1} representative diagrams for the correlation
function
$\langle\rho(E,\bm{r})\rho(E+\omega,\bm{r}+\bm{R})\rangle$.
Each LDOS is given by a fermionic loop dressed by interaction lines.
Averaging each loop over disorder generates diffusive
vertex corrections and yields the ZBA. On the other hand, diffusons connecting the
loops lead to multifractal correlations. The RG effectively sums up the leading
contributions of the diagrams with multiple interaction lines and
intra- and inter-loop diffusons inserted in all possible ways.

It is instructive to compare our findings with the known results for the
Anderson transition in $d=2+\epsilon$ dimensions in the absence of interactions.
In the case of non-interacting disordered fermions without time reversal
symmetry (the Wigner-Dyson unitary class A), the
$\beta$-function, the critical point and the localization length exponent are
known
up to the five-loop order \cite{HikamiWegnerB}
\begin{equation}
-\frac{dt}{d\ln y} = \beta^{(0)}(t) = \epsilon t - \frac{1}{2} t^3 - \frac{3}{8} t^5 + O(t^6) ,
\label{eq2E4}
\end{equation}
$t_*=(2\epsilon)^{1/2}(1-3\epsilon/4)+O(\epsilon^{5/2})$, and
$\nu=1/2\epsilon-3/4+O(\epsilon)$. The anomalous dimensions of operators which
determine the scaling behavior of the LDOS moments have been computed at the
four-loop level \cite{Wegner} with the result
\begin{equation}
\zeta_q^{(0)}(t) = \frac{q(1-q) t}{2}\left ( 1+ \frac{3 t^2}{8} + \frac{3
\zeta(3)}{16} q(q-1) t^3\right )+ O(t^5) ,
\label{eq2E5}
\end{equation}
where $\zeta(3)\approx 1.2$ stands for the Riemann zeta function.
This leads to the following expression for the corresponding multifractal exponents:
\begin{equation}
\Delta_q^{(0)} = q(1-q) \left (\frac{\epsilon}{2}\right )^{1/2} - \frac{3 \zeta(3)}{32} q^2(q-1)^2 \epsilon^2  +  O(\epsilon^{5/2}) .
\label{eq2E6}
\end{equation}
Comparing Eqs. \eqref{eq2E1} and \eqref{eq2E4}, one sees that Coulomb
interaction changes the $\beta$-function and, consequently, the fixed point
and critical exponents. Thus, Anderson transitions
with and without Coulomb interaction belong to  different universality
classes.

While we have shown that multifractality of the LDOS persists in the presence
of
Coulomb interaction, the values of the multifractal dimensions,
Eq.\eqref{eq2E3}, are essentially different from their non-interacting
counterparts \eqref{eq2E6}. This happens both because of a difference in the
corresponding scaling functions [cf. Eqs. \eqref{eq2E2} and \eqref{eq2E5}]
and because of different values of critical resistance $t_*$.
We mention that in both cases in the two-loop approximation the
spectrum of anomalous dimensions $\Delta_q$ (and thus the so-called singularity
spectrum $f(\alpha)$ that is obtained by the Legendre transformation (see
e.g. Ref. \cite{Evers08}) is
parabolic, $\Delta_q \simeq \gamma q(1-q)$. It is expected, however, that
higher-loop contribution will break the exact parabolicity in the Coulomb case,
in analogy with what happens (in the four-loop order) in the non-interacting
model.

For small $\epsilon$, when the values of the exponents are parametrically
controlled, the Coulomb interaction considerably reduces numerical values of
the anomalous exponent, i.e. weakens multifractality. As an example, for
$\epsilon=1/8$ we get $\gamma=0.25$ in the absence and $\gamma=0.03$ in
the presence of interaction. In the physically most interesting case
of dimensionality $d=3$, i.e. $\epsilon=1$, we can only use the one-loop term as an
estimate. This yields in the non-interacting case  $\gamma=0.7$ and
$\eta=-\Delta_2=1.4$, in fairly well agreement with numerical results. An
analogous estimate based on our results for the interacting system yields
$\gamma=0.25$ and $\eta=-\Delta_2=0.5$. Note that at $\epsilon=1$
the second-loop term in Eq.~\eqref{eq2E3} is numerically of the same order
(by absolute value) as the one-loop term; thus, this estimate is expected
to be quite rough.

\begin{figure}[t]
\centerline{\includegraphics[width=40mm]{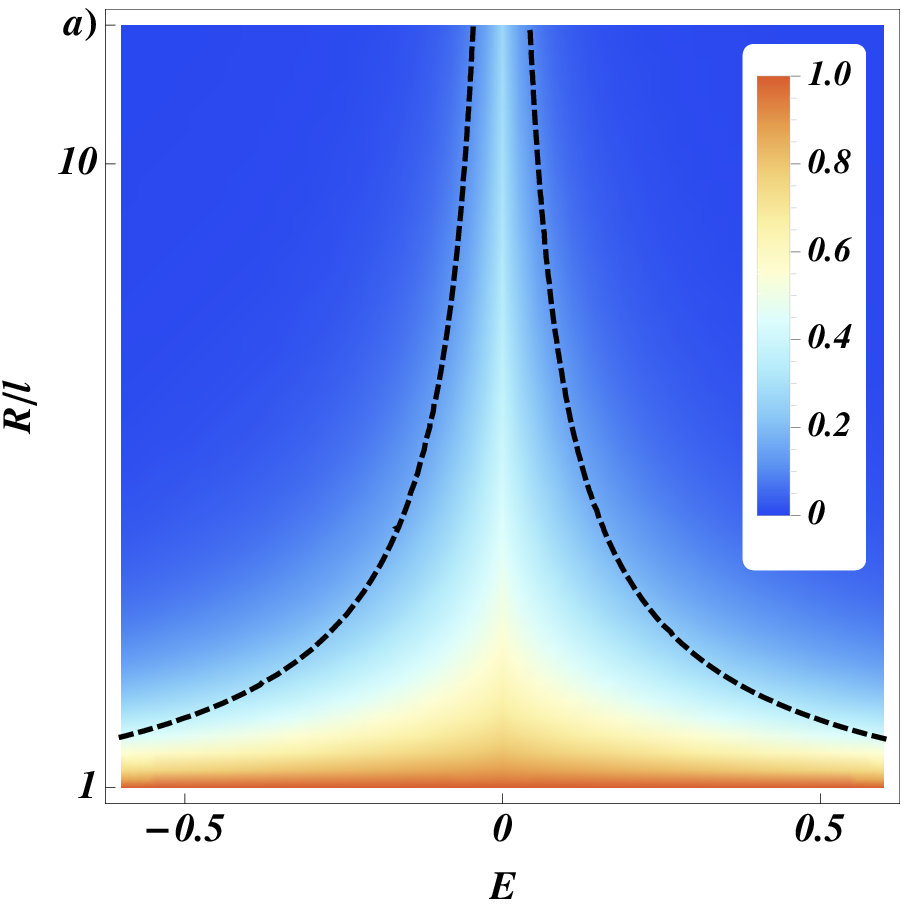}\hspace{.5cm}\includegraphics[width=40mm]{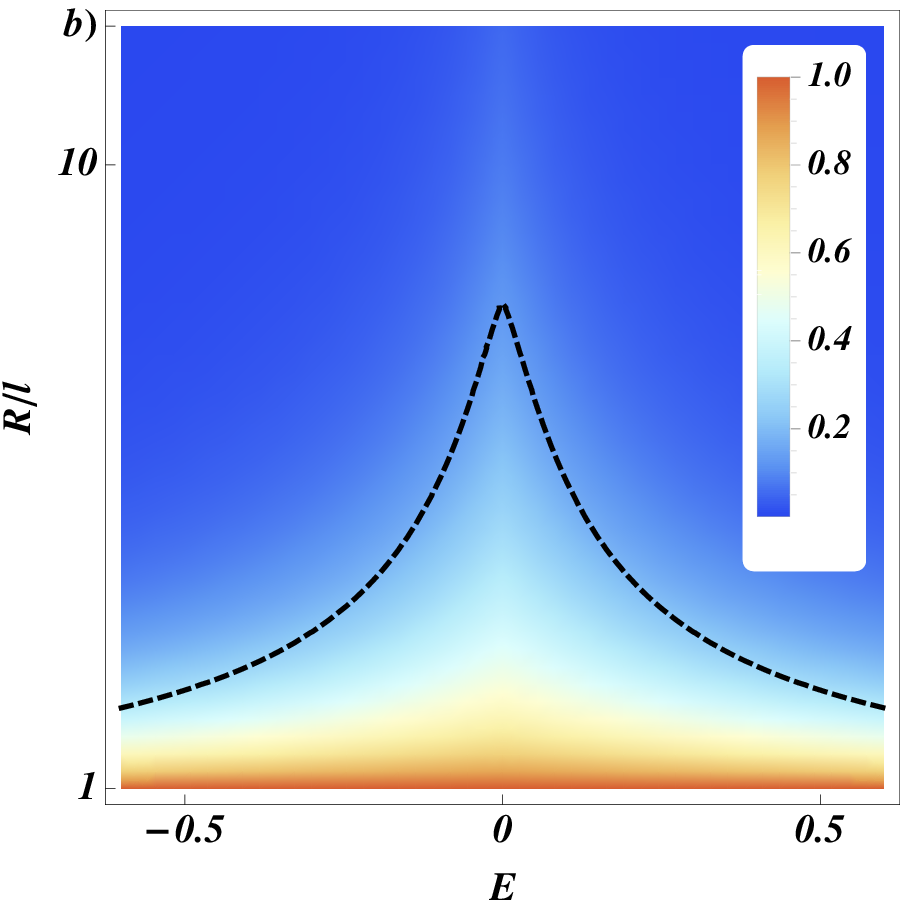}}
\caption{(Color online) Schematic color-code plot of the autocorrelation
function $\langle
[\rho(E,\bm{r})-\langle\rho(E)\rangle][\rho(E,\bm{r}+\bm{R})-\langle\rho(E)\rangle]
\rangle/\langle\rho(E)\rangle^2$ for the system a) at the critical point,
$g=g_*$, and b) slightly on the metallic side, $(g-g_*)/g_*=0.2$. The energy is
measured in units of elastic scattering rate which sets the ultraviolet
cutoff of the NL$\sigma$M theory. The dashed line in a) and b) corresponds to
$R/l\sim(|E|^{2/z}+|g/g_*-1|^{2\nu})^{-1/2}$. }
\label{Figure2a2b}
\end{figure}

To visualize the critical LDOS correlations near a
metal-insulator transition,
we present in Fig. \ref{Figure2a2b} a color-code plot of the autocorrelation
function
$\langle[\rho(E,\bm{r})-\langle\rho(E)\rangle][\rho(E,\bm{r}+\bm{R})
-\langle\rho(E)\rangle]\rangle/\langle\rho(E)\rangle^2$
[cf. Eq. \eqref{e5}]. This presentation is
analogous to Figs. 4A and 4B of the experimental
paper \cite{richardella10}. For this plot, we have chosen the following values
of the critical exponents: $\nu=1$, $z=1.5$, $\eta=0.5$, which are
theoretical estimates obtained by taking $\epsilon=1$ in the one-loop results
for the case of Coulomb interaction.
The left panel (Fig. \ref{Figure2a2b}a) corresponds to the case when the system is exactly at the
transition point, $g=g_*$. We see the long-range multifractal correlation at
low energies; since $\nu>1/z$ (as is also the case for experimental estimates
of the corresponding exponents at 3D metal-insulator transitions and at quantum
Hall transitions), the range of correlation $\mathcal{L}$ is controlled by the
dephasing length $L_\phi$. In the right panel (Fig. \ref{Figure2a2b}b), the system is slightly off the
transition, i.e., $g-g_*$ is now non-zero. In this case
$\mathcal{L}$ is governed by the correlation (localization) length $\xi$ in a
certain
window around zero energy and by $L_\phi$ outside this window. All essential
features of Fig. \ref{Figure2a2b} compare well with  Fig.4 of Ref.~\cite{richardella10}.

To summarize, we have shown that the multifractal fluctuations and correlations
of the LDOS persist in the presence of Coulomb interaction but the
spectrum of multifractal exponents is modified. By using the
NL$\sigma$M approach, we have calculated the
multifractality spectrum of an interacting system
without time-reversal and spin symmetries up to the two-loop
order in $2+\epsilon$ dimensions. Our results are in an overall
agreement with the experimental data of Ref.~\cite{richardella10}.

We hope that our work will motivate further experimental studies of
multifractality of interacting electrons near metal-insulator and quantum Hall
transitions. On the theoretical side, our paper paves a way to a systematic
investigation of multifractality at interacting critical points of localization
transitions. There exists by now a vast knowledge on properties of
multifractality in the absence of interaction, including, in particular, systems
of different symmetry classes and different dimensionalities, symmetries of
mutlifractal spectra, termination and freezing, implications of
conformal symmetry, connection to entanglement entropy, and manifestation of
multifractality in various observables
\cite{Evers08,jia08,obuse10,gruzberg11,gruzberg13}. In the presence of Coulomb
interaction, the corresponding physics remains to be explored.
In addition to metal-insulator transitions and transitions
between different phases of topological insulators, we envision a possibility
to extend this analysis also to superconductor-insulator transitions.

While we were preparing this paper for publication,  a preprint appeared
\cite{amini13} where an analogous problem is addressed numerically within
a self-consistent Hartree-Fock approximation.

We thank A.~Yazdani, A.W.W.~Ludwig, B.I.~Shklovskii, and M.A.~Skvortsov for discussions.
The work was supported by DFG-RFBR in the framework of SPP 1285, BMBF, Russian
President grant No. MK-4337.2013.02, Dynasty Foundation, RFBR grant No.
12-02-00579,
Russian Ministry of Education and Science under contract No. 8385,
and by RAS Programs ``Quantum mesoscopic and disordered  systems'',
``Quantum Physics of Condensed Matter'', and ``Fundamentals of nanotechnology
and nanomaterials''.


\newpage
\begin{widetext}

\centerline{\Large ONLINE SUPPORTING INFORMATION:}
\centerline{\Large Multifractality at Anderson transitions with Coulomb interaction }

\section{I.\, Nonlinear $\sigma$-model: Definitions}
\label{SM_SecI}

Low-energy physics of the system of disordered fermions with Coulomb interaction in the absence of time reversal and spin rotational symmetries (symmetry class denoted as ``MI(LR)'' in Ref.~[S1]) is described by the action of the nonlinear $\sigma$-model (NLSM). It is given as a sum of the non-interacting part, $S_\sigma$,
and the contribution arising from the interaction in the particle-hole singlet channel, $S_{\rm int}$:~[S1,S2]
\begin{gather}
S=S_\sigma + S_{\rm int},
\label{SsGen}
\end{gather}
where
\begin{align}
S_\sigma &= -\frac{g}{4} \int d\bm{r} \tr (\nabla Q)^2 + 4\pi T z \int d\bm{r} \tr \eta (Q-\Lambda) ,
\label{Ss} \\
S_{\rm int} & =- \pi T \Gamma_s \sum_{\alpha,n}
\int d\bm{r} \tr \Bigl [I_n^\alpha Q\Bigr ] \tr \Bigl [I_{-n}^\alpha Q\Bigr ] .
\label{Srho}
\end{align}
Here $g$ is the Drude conductivity (in units $e^2/h$) and we use the following matrices
\begin{gather}
\Lambda_{nm}^{\alpha\beta} = \sgn n\, \delta_{nm} \delta^{\alpha\beta}, \qquad
\eta_{nm}^{\alpha\beta}=n\, \delta_{nm}\delta^{\alpha\beta}, \qquad
(I_k^\gamma)_{nm}^{\alpha\beta}=\delta_{n-m,k}\delta^{\alpha\beta}\delta^{\alpha\gamma}
\end{gather}
with $\alpha, \beta, \gamma$ standing for replica indices and $n,m$ corresponding to
Matsubara fermionic energies $\varepsilon_n = \pi T (2n+1)$.
The matrix field $Q(\bm{r})$ (as well as the trace $\tr$) acts in the replica and Matsubara spaces. It obeys the following constraints:
\begin{gather}
Q^2=1, \qquad \tr Q = 0, \qquad Q^\dag =Q .
\end{gather}

We use notation similar to that of the paper by Baranov, Pruisken, and \v{S}kori\'{c} [S3].
However, in order to avoid notational confusion, it is instructive to compare our notation with that
of the reviews by Finkel'stein [S2] and by Belitz and
Kirkpatrick [S1].  The interaction term \eqref{Srho} coincides
with the term in Eq. (3.9a) of Ref.~[S2] with
 the coupling constant $\Gamma_s = -\pi \rho_0 Z/4$
and with the term in
Eq. (3.92d) in Ref.~[S1] with $\Gamma_s=K^{(1)}$.
Here $\rho_0$ is the thermodynamic density of states. Finally, the parameters $g$ and $z$ in Eq.~(\ref{Ss}) are related by $g = 4\pi \rho_0 D$ and $z= \pi \rho_0 Z/4$ to the corresponding parameters introduced in Ref.~[S2] and by $g=16/G$ and $z=H/2$ to those in Ref.~[S1]. We mention that Ref.~[S2] focuses on the case of unscreened (long-range) Coulomb interaction. Therefore, the singlet interaction amplitude $\Gamma_s$ is expressed through the frequency renormalization factor $Z$ there. In what follows, we consider a general case of an arbitrary range interaction for which these quantities are independent variables.

For the perturbative (in $1/g$) expansion we shall use the square-root parametrization
\begin{gather}
Q = W +\Lambda \sqrt{1-W^2}, \qquad W= \begin{pmatrix}
0 & w\\
\bar{w} & 0
\end{pmatrix} .
\label{Qparam}
\end{gather}
We adopt the following notations: $W^{\alpha\beta}_{n_1n_2} = w^{\alpha\beta}_{n_1n_2}$ and $W^{\alpha\beta}_{n_2n_1} = \bar{w}^{\alpha\beta}_{n_2n_1}$ with $n_1\geqslant 0$ and $n_2< 0$. The elements  $w^{\alpha\beta}_{n_1n_2}$ and $\bar{w}^{\alpha\beta}_{n_2n_1}$ are independent real variables.

For the purpose of regularization in the infrared, it is convenient to add the following term to the NLSM action 
\eqref{SsGen}:
\begin{equation}
S \to S + \frac{g h^2}{4} \int d \bm{r}\tr \Lambda Q .
\label{SsGenFull}
\end{equation}
Expanding the NLSM action \eqref{SsGenFull} to the second order in $W$, we find
\begin{align}
S^{(2)} & =-\frac{g}{4} \sum_{n_1n_2}\sum_{\alpha\beta}
\int \frac{d\bm{q}}{(2\pi)^d} \Bigl [q^2+h^2+ \frac{8\pi Tz}{g}n_{12} \Bigr]
w^{\alpha\beta}_{n_1n_2}(\bm{q})\bar{w}^{\beta\alpha}_{n_2n_1}(-\bm{q})  \notag \\
& - 2\pi T \Gamma_s \sum_{n_1n_2n_3n_4}\sum_{\alpha} \delta_{n_{12},n_{34}} 
\int \frac{d\bm{q}}{(2\pi)^d} 
w^{\alpha\alpha}_{n_1n_2}(\bm{q})\bar{w}^{\alpha\alpha}_{n_4n_3}(-\bm{q}) ,
\label{SsGauss}
\end{align}
where $n_{12} = n_1-n_2$. Hence, the propagator of $W$ fields becomes
\begin{gather}
\langle w^{\alpha_1\beta_1}_{n_1n_2}(\bm{q})\bar{w}^{\beta_2\alpha_2}_{n_4n_3}(-\bm{q})\rangle = \frac{4}{g} D_q(i\omega_{n_{12}}) \delta^{\alpha_1\alpha_2} \delta^{\beta_1\beta_2}
\left [\delta_{n_1n_3}\delta_{n_2n_4} - \frac{8\pi T\Gamma_s}{g} \delta^{\alpha_1\beta_1} \delta_{n_{12},n_{34}}D^s_q(i\omega_{n_{12}}) \right ],
\\
\Bigl [ D_q(i\omega_{n_{12}}) \Bigr ]^{-1} =q^2+\frac{4z}{g}\omega_{n_{12}}, \qquad 
\Bigl [D^s_q(i\omega_{n_{12}})\Bigr ]^{-1} =q^2+\frac{4(z+\Gamma_s)}{g}\omega_{n_{12}}
 . \label{Prop}
\end{gather}
Here $\omega_{n_{12}} = \varepsilon_{n_1}-\varepsilon_{n_2}$. Note that $\langle w w\rangle$ and $\langle \bar{w}\bar{w} \rangle$ are zero.

\section{II. \, The average of local density of states: One-loop approximation}

In this section, we remind known results for the local density of states (LDOS). It is related with the single-particle Green function as $\rho(E, \bm{r}) = (-1/\pi)\im G(E; \bm{r}, \bm{r})$. In the NLSM approach, the average LDOS $\langle \rho(E, \bm{r}) \rangle$ can be obtained after analytic continuation from Matsubara frequencies ($\varepsilon_{n_1}\to E+i0^+$) of the following function 
\begin{equation}
P_1(i\varepsilon_{n_1}) = \rho_0 \langle Q_{n_1n_1}^{\alpha\alpha}(\bm{r}) \rangle .
\end{equation}
Here $\alpha$ is a fixed replica index and $\rho_0$ is the single-particle density of states at energy of the order of inverse elastic scattering time $1/\tau$, playing a role of high-energy cutoff of the theory. By using Eqs \eqref{Qparam} and \eqref{Prop},  one finds
in the one-loop approximation
\begin{equation}
P_1(i\varepsilon_{n_1}) = 1 + \frac{16 \pi T \Gamma_s}{g^2} \sum_{\varepsilon_n> 0} \int \frac{d^d\bm{q}}{(2\pi)^d} D_q(i\varepsilon_{n_1}+i \varepsilon_n)D_q^s(i\varepsilon_{n_1}+i \varepsilon_n)  .
\label{eqPR1}
\end{equation}
After analytic continuation in Eq.~\eqref{eqPR1}, we obtain
\begin{equation}
\langle \rho(E) \rangle =  \rho_0 \Bigl [ 1 + \frac{4 \Gamma_s}{g^2}\Imag  \int_{-\infty}^\infty d\omega \tanh \frac{\omega-E}{2T} \int \frac{d^d\bm{q}}{(2\pi)^d} D^R_q(\omega)D_q^{s R}(\omega)\Bigr ] .\label{eqN1}
\end{equation}
Here $D^R_q(\omega)$ and $D_q^{s R}(\omega)$ are retarded propagators corresponding to Matsubara propagators $D_q(i\omega_m)$ and $D_q^s(i\omega_m)$, respectively.
One can check that Eq.~\eqref{eqN1} reproduces the well-known perturbative result for the zero-bias anomaly (ZBA) [S4].  

To simplify analysis, it is convenient to set temperature $T$ and energy $E$ to zero and study dependence of $\langle \rho\rangle$ on the infrared regulator $h^2$. Then, in $d=2+\epsilon$ dimensions, we find
\begin{equation}
\langle \rho \rangle = \rho_0 Z^{1/2}, \qquad  Z = 1 - 2 t \ln(1+\gamma_s)\frac{h^\epsilon}{\epsilon} +O(\epsilon) .
\label{eqAvDOS}
\end{equation}
Here $\gamma_s=\Gamma_s/z$, and $t = 4 \Omega_d/g$ where $\Omega_d= S_d/[2(2\pi^d)]$ ($S_d=2\pi^{d/2}/\Gamma(d/2)$ is the area of the $d$-dimensional sphere). We notice the well-known peculiarity of the case of Coulomb interaction ($\gamma_s=-1$) for which the formally divergent term $ \ln(1+\gamma_s)$ in Eq. \eqref{eqAvDOS} emerges in addition to $1/\epsilon$ factor. We emphasize that the terms divergent at $\gamma_s\to -1$ never appear in renormalization of gauge-invariant quantities, e.g. conductivity.

\section{III.\, The second moment of the LDOS}

\subsection{A.\, Two-point irreducible correlation function of the LDOS}

Let us define the irreducible two-point correlation function of the LDOS
\begin{equation}
K_2(E,\bm{r};E^\prime, \bm{r^\prime}) = \langle \rho(E, \bm{r})  \rho(E^\prime, \bm{r^\prime}) \rangle - \langle \rho(E, \bm{r}) \rangle \langle \rho(E^\prime, \bm{r^\prime}) \rangle .
\end{equation}
Since we are mainly interested in the second moment of the LDOS, in what follows we consider $K_2(E,\bm{r};E^\prime, \bm{r^\prime})$ at coinciding spatial points only. It can be obtained from the function
 \begin{equation}
 K_2 = \frac{\rho_0^2}{2}\re \Bigl (P_2^{\alpha_1\alpha_2}(i\varepsilon_{n_1},i\varepsilon_{n_3}) - P_2^{\alpha_1\alpha_2}(i\varepsilon_{n_1},i\varepsilon_{n_4})) \Bigl ) 
 \label{eqK2def0}
 \end{equation}
after analytic continuation to the real frequencies:  $\varepsilon_{n_1} \to E+i0^+$, $\varepsilon_{n_3} \to E^\prime+i0^+$, and $\varepsilon_{n_4} \to E^\prime-i0^+$. Here 
 \begin{gather}
 P_2^{\alpha_1\alpha_2}(i\varepsilon_{n},i\varepsilon_{m}) = \langle\langle  Q_{nn}^{\alpha_1\alpha_1}(\bm{r}) \cdot Q_{mm}^{\alpha_2\alpha_2}(\bm{r}) \rangle \rangle - \langle Q_{nm}^{\alpha_1\alpha_2}(\bm{r}) Q_{mn}^{\alpha_2\alpha_1}(\bm{r}) \rangle 
 ,
 \label{eqP2corr}
 \end{gather}
where $\langle \langle A \cdot B\rangle \rangle =  \langle A B\rangle -  \langle A \rangle \langle B\rangle$. Replica indices $\alpha_1$ and $\alpha_2$ are different in Eq. \eqref{eqK2def0}: $\alpha_1\neq \alpha_2$ such that the two-point correlation function $K_2$ measures {\it mesoscopic} fluctuations of the LDOS.
 
\subsection{B.\, One-loop result for $K_2$}

In the one-loop approximation, one finds
\begin{gather}
[P_2^{\alpha_1\alpha_2}]^{(1)}(i\varepsilon_{n_1},i\varepsilon_{n_3}) = 0, \qquad  
[P_2^{\alpha_1\alpha_2}]^{(1)}(i\varepsilon_{n_1},i\varepsilon_{n_4}) = - \langle w^{\alpha_1\alpha_2}_{n_1n_4} \bar{w}^{\alpha_2\alpha_1}_{n_4n_1}\rangle_0 = - \frac{4}{g} \int_q D_q(i\omega_{n_{14}})
,  
\end{gather}
where $\int_q \equiv \int d^d \bm{q}/(2\pi)^d$, $\langle \dots \rangle_0$ denotes average with respect to the quadratic part \eqref{SsGauss} of the NLSM action. Hence,
we obtain
\begin{equation}
K_2^{(1)}(E,\bm{r};E^\prime, \bm{r})  = \rho_0^2\,\frac{2}{g} \Real \int_q D^R_q(E-E^\prime)  . \label{eq1loopK2_2}
\end{equation}
Setting $E=E^\prime$ and using $h^2$ as the infrared regulator, one finds
\begin{equation}
K_2^{(1)}  =  - \rho_0^2\, t \frac{h^\epsilon}{\epsilon} +O(\epsilon) . \label{eq1loopK2_3}
\end{equation}

\subsection{C.\, Two-loop result for $K_2$}

We start evaluation of the two-loop contribution to the irreducible two-point correlation function $K_2$ from $P^{\alpha_1\alpha_2}_2(i\varepsilon_{n_1},i\varepsilon_{n_3})$. In the two-loop approximation, one needs to take into account only terms with four $W$: 
\begin{equation}
[P_2^{\alpha_1\alpha_2}]^{(2)}(i\varepsilon_{n_1},i\varepsilon_{n_3}) = \frac{1}{4} \sum_{n_6n_8}\sum_{\beta_1\beta_2} \Bigl [ \langle \langle 
w^{\alpha_1\beta_1}_{n_1n_6}\bar{w}^{\beta_1\alpha_1}_{n_6n_1} \cdot
w^{\alpha_2\beta_2}_{n_3n_8}\bar{w}^{\beta_2\alpha_2}_{n_8n_3}  
\rangle \rangle_0 - 
\langle 
w^{\alpha_1\beta_1}_{n_1n6}\bar{w}^{\beta_1\alpha_2}_{n_6n_3} 
w^{\alpha_2\beta_2}_{n_3n_8}\bar{w}^{\beta_2\alpha_1}_{n_8n_1}  
\rangle_0 \Bigr ] .
\end{equation}
By using Wick theorem and Eq.~\eqref{Prop}, we find
\begin{equation}
[P_2^{\alpha_1\alpha_2}]^{(2)}(i\varepsilon_{n_1},i\varepsilon_{n_3})  = 
 \frac{32\pi T \Gamma_s}{g^3} \sum_{\varepsilon_n>0} \int_{q,p}D_q(i\varepsilon_{n_1}+i \varepsilon_n)  \Bigl [
D_p(i\varepsilon_{n_3}+i \varepsilon_n)D^s_p(i\varepsilon_{n_3}+i \varepsilon_n)+D^s_q(i\varepsilon_{n_1}+i \varepsilon_n)
D_p(i\varepsilon_{n_3}+i \varepsilon_n)  \Bigr ] .
\label{eq2loopK2_P2++1}
\end{equation} 
Performing analytical continuation, we obtain
\begin{equation}
[P_2^{\alpha_1\alpha_2}]^{(2)}(E,E^\prime)  = 
 \frac{8 \Gamma_s}{i g^3} \int_{-\infty}^\infty d\omega \tanh\frac{\omega}{2T} \int_{q,p}\Bigl [
D^R_q(\omega+E) D^R_p(\omega+E^\prime)D^{s R}_p(\omega+E^\prime)+
D^R_q(\omega+E^\prime) D^R_p(\omega+E)D^{sR}_p(\omega+E) \Bigr ] .
\label{eq2loopK2_P2++11}
\end{equation} 
Setting $E=E^\prime=T=0$, one can derive
\begin{equation}
[P_2^{\alpha_1\alpha_2}]^{(2)}(E,E^\prime) \to \frac{t^2\,h^{2\epsilon}}{\epsilon^2}\Bigl [
\ln (1+\gamma_s) -\frac{\epsilon}{4} \ln^2(1+\gamma_s)\Bigr ]+O(1).
\label{eq2loopK2_P2++2}
\end{equation}

The two-loop contribution to $P^{\alpha_1\alpha_2}_2(i\varepsilon_{n_1},i\varepsilon_{n_4})$ can be written as
\begin{equation}
[P^{\alpha_1\alpha_2}_2]^{(2)}(i\varepsilon_{n_1},i\varepsilon_{n_4}) = 
-\frac{1}{4} \sum_{n_5n_6}\sum_{\beta_1\beta_2} \langle \langle 
w^{\alpha_1\beta_1}_{n_1n_6}\bar{w}^{\beta_1\alpha_1}_{n_6n_1} \cdot
\bar{w}^{\alpha_2\beta_2}_{n_4n_5}{w}^{\beta_2\alpha_2}_{n_5n_4}  
\rangle \rangle_0 - \Bigl \langle w^{\alpha_1\alpha_2}_{n_1n_4} \bar{w}^{\alpha_2\alpha_1}_{n_4n_1}
\Bigl [ S^{(4)}_0+S^{(4)}_{\rm int}+\frac{1}{2} \left (S^{(3)}_{\rm int}\right )^2\Bigr ] \Bigr \rangle_0  .
\label{eqP2+-0}
\end{equation} 
Here the term
\begin{align}
S^{(4)}_0  & = \frac{g}{32} \int_{q_1}\int_{q_2} \int_{q_3} \int_{q_4} \delta(\bm{q_1}+\bm{q_2}+\bm{q_3}+\bm{q_4}) 
\sum_{\beta_1\beta_2\beta_3\beta_4}\sum_{n_5n_6n_7n_8}w^{\beta_1\beta_2}_{n_5n_6} \bar{w}^{\beta_2\beta_3}_{n_6n_7}
w^{\beta_3\beta_4}_{n_7n_8} \bar{w}^{\beta_4\beta_1}_{n_8n_5}
 \notag \\
 & \hspace{3cm} \times 
\Bigl [ (\bm{q_1}+\bm{q_2})(\bm{q_3}+\bm{q_4})+(\bm{q_1}+\bm{q_4})(\bm{q_2}+\bm{q_3})  -2h^2 - \frac{8\pi T z}{g} (n_{56}+n_{78})\Bigr ] , 
\end{align}
appears in the expansion of $S_\sigma$ and the regulator term to the forth order in $W$. The expansion of the singlet interaction term $S_{\rm int}$ results in the following third and forth order terms:
\begin{equation}
S^{(3)}_{\rm int}  = \pi T \Gamma_s \sum_{\beta,n} \int d\bm{r} \tr I^{\beta}_{n} W
\tr I^{\beta}_{-n}\Lambda W^2 ,\qquad 
S^{(4)}_{\rm int} = -\frac{\pi T \Gamma_s}{4} \sum_{\beta,n} \int d\bm{r} \tr I^{\beta}_{n} \Lambda W^2
\tr I^{\beta}_{-n}\Lambda W^2  .
\end{equation} 
After evaluation of averages in Eq.~\eqref{eqP2+-0}, we find
\begin{align}
[P_2^{\alpha_1\alpha_2}]^{(2)}(i\varepsilon_{n_1},i\varepsilon_{n_4}) & =  - \left [ \frac{2}{g} \int_q D_q(i\omega_{n_{14}})\right ]^2 - \frac{32 \pi T \Gamma_s}{g^3} \int_{q,p} \sum_{\varepsilon_n > 0} \Bigl [ D^2_q
(i\omega_{n_{14}}) D_p^s(i\varepsilon_{n}+i\varepsilon_{n_1})+D^2_q
(i\omega_{n_{14}}) D_p^s(i\varepsilon_{n}-i\varepsilon_{n_4}) \notag \\
& + D_q
(i\omega_{n_{14}}) D_p(i\varepsilon_{n}+i\varepsilon_{n_1})D_p^s(i\varepsilon_{n}+i\varepsilon_{n_1}) +D_q
(i\omega_{n_{14}}) D_p(i\varepsilon_{n}-i\varepsilon_{n_4})D_p^s(i\varepsilon_{n}-i\varepsilon_{n_4})\Bigr ] 
\notag \\ 
&  + \frac{64 \pi T \Gamma_s}{g^3} \int_{q,p} \sum_{\omega_n>0} D^2_q (i\omega_{n_{14}}) D_p (i\omega_{n_{14}}+i\omega_n)
\Bigl [1 - \frac{4 \Gamma_s \omega_n }{g} D_{\bm{q}+\bm{p}}^s
(i\omega_{n})  \Bigr ] \notag\\
&  +  \frac{32 \pi T \Gamma_s}{g^3} \int_{q,p} \sum_{\epsilon_{n_1}>\omega_n>0}
D^{2}_p (i\omega_{n_{14}})\Bigl [ 1 - \frac{4 \Gamma_s \omega_n }{g} D_{\bm{q}+\bm{p}}^s
(i\omega_{n})  \Bigr ]
  D_q (i\omega_{n_{14}}-i\omega_n) \notag \\
  &  +  \frac{32 \pi T \Gamma_s}{g^3} \int_{q,p} \sum_{-\epsilon_{n_4}>\omega_n>0}
D^{2}_p (i\omega_{n_{14}})\Bigl [ 1 - \frac{4\Gamma_s \omega_n }{g} D_{\bm{q}+\bm{p}}^s
(i\omega_{n})  \Bigr ]
D_q (i\omega_{n_{14}}-i\omega_n) 
.
\label{eq2loopK2_P2+-1}
\end{align}
Performing analytic continuation to the real frequencies in Eq. \eqref{eq2loopK2_P2+-1}, we obtain
\begin{align}
[P_2^{\alpha_1\alpha_2}]^{(2)}(E,E^\prime) & =  - \left [ \frac{2}{g} \int_q D^R_q(E-E^\prime)\right ]^2 
 - \frac{8 \Gamma_s}{i g^3} \int_{q,p} D^R_q(E-E^\prime) \int\limits_{-\infty}^\infty d\omega \tanh\frac{\omega}{2T}  \Bigl [ D^R_q(E-E^\prime) D_p^{sR}(\omega+E)
 \notag \\
 & +D^R_q
(E-E^\prime) D_p^{sR}(\omega-E^\prime)  + D_p^R(\omega+E)D_p^{sR}(\omega+E) + D^R_p(\omega-E^\prime)D_p^{sR}(\omega-E^\prime)\Bigr ] 
\notag \\ 
&  + \frac{16 \Gamma_s}{i g^3} \int_{q,p} [D^R_q (E-E^\prime)]^2  \int\limits_{-\infty}^\infty d\omega \coth\frac{\omega}{2T}D^R_p (\omega+E-E^\prime)
\Bigl [1 + \frac{4 i\Gamma_s \omega}{g} D_{\bm{q}+\bm{p}}^{sR}
(\omega)  \Bigr ] \notag \\
&+  \frac{8 \Gamma_s}{i g^3} \int_{q,p} [D^R_q (E-E^\prime)]^2 \int\limits_{-\infty}^\infty d\omega D^R_p (E-E^\prime-\omega) 
\Bigl [ 1 + \frac{4 i \Gamma_s \omega}{g} D_{\bm{q}+\bm{p}}^{s R}
(\omega)  \Bigr ] \notag \\
& \hspace{2cm} \times \Bigl [ 2 \coth\frac{\omega}{2T} -\tanh\frac{\omega-E}{2T}-\tanh\frac{\omega+E^\prime}{2T}\Bigr ]
.
\label{eq2loopK2_P2+-1a}
\end{align}
Again setting $E=E^\prime=T=0$,  one can derive
\begin{equation}
[P_2^{\alpha_1\alpha_2}]^{(2)}(E,E^\prime) \to - \frac{t^2\,h^{2\epsilon}}{\epsilon^2}\Bigl [
1+ 3 \ln (1+\gamma_s) + 2 f(\gamma_s) -
\epsilon\frac{2+\gamma_s}{\gamma_s}  \Bigl ( \ln(1+\gamma_s) + \liq(-\gamma_s) + \frac{1}{4}\ln^2(1+\gamma_s)\Bigr ) \Bigr ] +O(1) ,
\label{eq2loopK2_P2+-2}
\end{equation}
where $f(x) = 1 - (1+1/x)\ln(1+x)$. Combining together Eqs \eqref{eq2loopK2_P2++2} and \eqref{eq2loopK2_P2+-2}, we obtain the following two-loop  
contribution to the irreducible two-point correlation function:
\begin{equation}
K_2^{(2)}  = \rho_0^2 \frac{t^2 h^{2\epsilon}}{\epsilon^2} \Biggl \{ \Bigl [ \frac{1}{2} + 2 \ln(1+\gamma_s) + f(\gamma_s) \Bigr ] + \epsilon \Bigl [ \frac{1}{2}\ln(1+\gamma_s) + f(\gamma_s) -1 -\frac{2+\gamma_2}{2\gamma_s} \liq(-\gamma_s) - \frac{1+\gamma_s}{4\gamma_s} \ln^2(1+\gamma_s)\Bigr ]\Biggr \}+O(1) .\label{eqK1_0}
\end{equation}

\subsection{D.\, Two-loop renormalization of the second moment of the LDOS} 
 
By using Eqs \eqref{eq1loopK2_3} and \eqref{eqK1_0}, we write the two-loop result for the irreducible two-point correlation function as follows
\begin{equation}
K_2  = \rho_0^2 \Biggl [-\frac{t\, h^\epsilon}{\epsilon} + \frac{t^2 h^{2\epsilon}}{\epsilon^2} \Biggl \{ \Bigl [ \frac{1}{2} + 2 \ln(1+\gamma_s) + f(\gamma_s) \Bigr ] + \epsilon \Bigl [ \frac{1}{2}\ln(1+\gamma_s) + f(\gamma_s) -1 -\frac{2+\gamma_2}{2\gamma_s} \liq(-\gamma_s) - \frac{1+\gamma_s}{4\gamma_s} \ln^2(1+\gamma_s)\Bigr ] \Biggr \} \Biggr ] . \label{eqK1ren}
\end{equation}

It is known [S3] that the momentum scale $h$ acquires renormalization. The corresponding renormalized momentum scale  $h^\prime$ can be defined as 
\begin{equation}
g^\prime h^{\prime 2}  \tr \Lambda^2 = g h^2 \langle \tr \Lambda Q \rangle , 
\end{equation}  
where $g^\prime$ denotes renormalized conductivity at the momentum scale $h^\prime$. In the one-loop approximation, one can find [S3]
\begin{equation}
h^\prime = h \Bigl [ 1 - \frac{t\,  h^\epsilon}{\epsilon} \Bigl (f(\gamma_s)+\frac{1}{2}\ln(1+\gamma_s) \Bigr )+O(\epsilon) \Bigr ] 
\label{eqhren}
\end{equation}
and [S5,S6,S7] 
\begin{equation}
g^\prime = g \Bigl [1+\frac{a_1 t\, h^\epsilon}{\epsilon} +O(\epsilon)\Bigr ], \qquad a_1 =2  f(\gamma_s) . 
\label{eqS1}
\end{equation}
We mention that 
\begin{equation}
g^\prime h^{\prime 2} = g h^2 Z^{1/2}
\end{equation}
as expected (see Ref. [S1]).  

By using Eq. \eqref{eqhren}, we can write the second moment of the LDOS in terms of the renormalized momentum scale $h^\prime$  and factor $Z$ as follows:
\begin{equation}
\langle \rho^2\rangle = Z\rho_0^2 +K_2 = \rho_0^2 Z m^\prime_2, \qquad m^\prime_2 = m_2 \Bigl [ 1+ \frac{b^{(2)}_1 t \, h^{\prime \epsilon}}{\epsilon}+ \frac{t^2 h^{\prime 2\epsilon}}{\epsilon^2} \Bigl (b^{(2)}_2+\epsilon b^{(2)}_3 +O(\epsilon^2) \Bigr )\Bigr ] ,
\label{eqM2def}
\end{equation}
where $m_2=1$ and
\begin{equation}
b^{(2)}_1 = -1, \qquad b^{(2)}_2 = \frac{1}{2}+f(\gamma_s), \qquad b^{(2)}_3 = -1 -\frac{2+\gamma_s}{2\gamma_s} \liq(-\gamma_s) - \frac{1+\gamma_s}{4\gamma_s} \ln^2(1+\gamma_s) .
\end{equation}
Here $\liq(x) = \sum_{k=1}^\infty x^k/k^2$ denotes dilogarithm. We remind that the interaction parameter $\gamma_s$ undergoes renormalization. At the one-loop level it is as follows [S7]: 
\begin{equation}
\gamma_s^\prime = \gamma_s \Bigl [ 1+ (1+\gamma_s) t\frac{h^{\prime \epsilon}}{\epsilon}\Bigr ] .
\label{eqGammaRen}
\end{equation}

In order to find the anomalous dimension of $m^\prime_2$, we introduce dimensionless quantity $\bar{t} = t^\prime h^{\prime \epsilon}$ and, using Eqs \eqref{eqS1}, \eqref{eqM2def}, \eqref{eqGammaRen}, express $t$, $\gamma_s$ and $m_2$  as
\begin{equation}
t =  (h^\prime)^{-\epsilon} \bar{t} Z_t(\bar{t},\gamma^\prime_s),\qquad \gamma_s = \gamma_s^\prime Z_\gamma(\bar{t},\gamma^\prime_s), \qquad m_2= m_2^\prime 
Z_{m_2}(\bar{t},\gamma^\prime_s) .
\end{equation}
To the lowest orders in $\bar{t}$ the renormalization parameters become
\begin{equation}
Z_t= 1 + \frac{a_1(\gamma^\prime_s)}{\epsilon}\bar{t}, 
\qquad Z_\gamma^{-1} = 1+ (1+\gamma_s) \frac{\bar{t}}{\epsilon}
\qquad 
Z_{m_2}^{-1} =  1 + \frac{b^{(2)}_1}{\epsilon}\bar{t} +
\frac{\bar{t}^2}{\epsilon^2} \Bigl [ b^{(2)}_2(\gamma_s^\prime) + b^{(2)}_1 a_1(\gamma_s^\prime) + \epsilon b^{(2)}_3(\gamma_s^\prime) \Bigr ] .
\label{eqZZZ}
\end{equation}
Here we use the fact that $b^{(2)}_1$ is independent of $\gamma_s$. Now the renormalization group functions can be derived in a standard manner from the conditions that $t$, $\gamma_s$ and $m_2$ does
not depend on the momentum scale $h^\prime$. We reproduce known one-loop results for the renormalization of the dimensionless resistance  and interaction parameter [S7]
\begin{equation}
-\frac{dt}{d\ln y} = \beta(t,\gamma_s) = \epsilon t - 2 f(\gamma_s) t^2 + O(t^3), \qquad 
-\frac{d\gamma_s}{d\ln y} = \beta_\gamma(t,\gamma_s) = \gamma_s(1+\gamma_s) t + O(t^2)  ,
\label{betafunct}
\end{equation}
and obtain two-loop result for the anomalous dimension of $m_2$
\begin{equation}
-\frac{d \ln m_2}{d \ln y} = \zeta_2(t,\gamma_s) =  -t - c(\gamma_s) t^2 + O(t^3) , \qquad c(\gamma_s) = 2 + \frac{2+\gamma_s}{\gamma_s} \liq(-\gamma_s) + \frac{1+\gamma_s}{2\gamma_s} \ln^2(1+\gamma_s) .
\label{eqm2RG}
\end{equation}
Here $y=1/h^\prime$ is the renormalization group running length scale and we omit `prime' and `bar' signs for a brevity. It is worthwhile to mention that $c(0) = 0$ as it is known for free electrons [S8], and $c(-1) = 2 -\pi^2/6\approx 0.36$. We emphasize that the relation $b^{(2)}_2 = b^{(2)}_1(b^{(2)}_1-a_1)/2$ guaranties the renormalizability of $m_2$, i.e. the absence in Eq.~\eqref{eqm2RG} of terms divergent in the limit $\epsilon\to 0$.

Therefore, the results of this section implies that the second moment of the LDOS at $E=T=0$ can be written as 
\begin{equation}
\langle \rho^2 \rangle = \langle \rho \rangle^2\, m_2  ,
\label{eqK3}
\end{equation}
where behavior of $m_2$ is governed by Eq.~\eqref{eqm2RG}. We mention that interaction affects the anomalous dimension of $m_2$ only at the two-loop level.

\section{IV.\, The $q$-th moment of the LDOS}

\subsection{A.\, General arguments}

In this section we demonstrate that in the two-loop approximation the $q$-th moment of the LDOS at $E=T=0$ can be written as 
\begin{equation}
\left \langle \rho^q\right \rangle =\langle \rho\rangle^q m_q  ,
\label{eqKmq}
\end{equation}
where the behavior of $m_q$ is determined by the following renormalization group equation:
\begin{equation}
-\frac{d \ln m_q}{d \ln y} = \zeta_q(t,\gamma_s) = \frac{q(1-q)}{2} t + \frac{q(1-q)}{2} c(\gamma_s) t^2 + O(t^3) .
\label{eqmqRG}
\end{equation}
Here the function $c(\gamma_s)$ is given in Eq. \eqref{eqm2RG}. We mention that Eq. \eqref{eqmqRG} implies 
\begin{equation}
m^\prime_q =  m_q \Bigl [ 1+ \frac{b_1^{(q)} t \, h^{\prime \epsilon}}{\epsilon}+ \frac{t^2 h^{\prime 2\epsilon}}{\epsilon^2} \Bigl [b^{(q)}_2+\epsilon b^{(q)}_3+O(\epsilon^2) \Bigr ]\Bigr ] 
\label{eqmqe}
\end{equation}
with $m_q=1$ and
\begin{gather}
b_1^{(q)}= \frac{q(q-1)}{2} b_1^{(2)}, \qquad  b_2^{(q)}= \frac{b_1^{(q)}}{2} (b_1^{(q)} - a_1)= 
\frac{1}{2}\Bigl [\frac{q(q-1)}{2}b_1^{(2)} \Bigr ]^2  -\frac{q(q-1)}{4}b_1^{(2)} a_1, \qquad b_3^{(q)}= \frac{q(q-1)}{2} b_3^{(2)} .
\label{eqmqe1}
\end{gather}
Let us consider the  irreducible $q$-th moment of the LDOS (with $q\geqslant 3$)
\begin{equation}
K_q = \bigl \langle \left ( \rho - \langle \rho \rangle \right )^q \bigr \rangle .
\end{equation}
Then we can write
\begin{equation}
\left \langle \rho^q \right \rangle  = \sum\limits_{j=0}^{q-1} (-1)^{q-1-j} C^q_j  \langle \rho^j\rangle \langle \rho \rangle^{q-j} + K_q .
\end{equation}
Provided Eqs \eqref{eqKmq} and \eqref{eqmqe} hold for all $0\leqslant j\leqslant q-1$, we find 
\begin{equation}
\left \langle \rho^q \right \rangle =
Z^q \tilde{m}_q + K_q, \qquad 
\tilde{m}_q =  m_q\Bigl [ 1+ \frac{\tilde{b}_1^{(q)} t \, h^{\prime \epsilon}}{\epsilon}+ \frac{t^2 h^{\prime 2\epsilon}}{\epsilon^2} \Bigl [\tilde{b}^{(q)}_2+\epsilon \tilde{b}^{(q)}_3 +O(\epsilon^2)\Bigr ]\Bigr ] ,
\label{eqmqe3}
\end{equation}
where 
\begin{gather}
\tilde{b}_1^{(q)}= \frac{k_q}{2} b_1^{(2)}, \qquad  \tilde{b}_2^{(q)}= 
\frac{l_q}{8}\Bigl [b_1^{(2)} \Bigr ]^2  -\frac{k_q}{4}b_1^{(2)} a_1, \qquad \tilde{b}_3^{(q)}= \frac{k_q}{2} b_3^{(2)}, \notag \\
k_q = \partial_x^2 \Bigl [x^q-(x-1)^q \Bigr ]\Biggl |_{x=1} , \qquad 
l_q = \partial_x^2\Bigl ( x^2 \partial_x^2 \Bigl [x^q-(x-1)^q \Bigr ]\Bigr )\Biggl |_{x=1}  .
\label{eqmqe4}
\end{gather}
As one can check, $k_q = q(q-1)$ for $q\geqslant 3$ whereas $l_q=k_q^2$ for $q\geqslant 5$. For $q=3$ and $q=4$ one finds $l_3=12$ and $l_4=120$. Since expression for $K_q$ involves connected contributions from averages of the number $q$ of matrices $Q$, there is no two-loop contribution to $K_q$ for $q\geqslant 5$. Therefore, with the help of Eq \eqref{eqmqe4}, we obtain result \eqref{eqmqe} for $q\geqslant 5$. The cases $q=3$ and $q=4$ need special consideration.

\subsection{B.\, The third irreducible moment of the LDOS}

The third irreducible moment $K_3$ can be obtained from the function
 \begin{align}
 K_3 &= \frac{\rho_0^3}{8} \Bigl (P_3^{\alpha_1\alpha_2\alpha_3}(i\varepsilon_{n_1},i\varepsilon_{n_3},i\varepsilon_{n_5}) - P^{\alpha_1\alpha_2\alpha_3}_3(i\varepsilon_{n_1},i\varepsilon_{n_3},i\varepsilon_{n_2}) - P^{\alpha_1\alpha_2\alpha_3}_3(i\varepsilon_{n_1},i\varepsilon_{n_2},i\varepsilon_{n_3})
 - P^{\alpha_1\alpha_2\alpha_3}_3(i\varepsilon_{n_2},i\varepsilon_{n_1},i\varepsilon_{n_3})\notag \\
  & + P^{\alpha_1\alpha_2\alpha_3}_3(i\varepsilon_{n_2},i\varepsilon_{n_4},i\varepsilon_{n_1}) + P^{\alpha_1\alpha_2\alpha_3}_3(i\varepsilon_{n_2},i\varepsilon_{n_1},i\varepsilon_{n_4})
 + P^{\alpha_1\alpha_2\alpha_3}_3(i\varepsilon_{n_1},i\varepsilon_{n_2},i\varepsilon_{n_4})-P^{\alpha_1\alpha_2\alpha_3}_3(i\varepsilon_{n_2},i\varepsilon_{n_4},i\varepsilon_{n_6})
 \Bigl ) 
  \end{align}
 after analytic continuation to the real frequencies: $\varepsilon_{n_{1,3,5}} \to E+i0^+$ and $\varepsilon_{n_{2,4,6}} \to E-i0^+$. Here
 \begin{align}
 P_3^{\alpha_1\alpha_2\alpha_3}(i\varepsilon_{n}, i\varepsilon_{m}, i\varepsilon_{k}) & = \langle Q_{nn}^{\alpha_1\alpha_1}(\bm{r}) Q_{mm}^{\alpha_2\alpha_2}(\bm{r}) Q_{kk}^{\alpha_3\alpha_3}(\bm{r}) \rangle - 3 \langle Q_{nn}^{\alpha_1\alpha_1}(\bm{r}) \rangle \langle  Q_{mm}^{\alpha_2\alpha_2}(\bm{r}) Q_{kk}^{\alpha_3\alpha_3}(\bm{r}) \rangle 
 \notag \\
 & -3 \langle Q_{nn}^{\alpha_1\alpha_1}(\bm{r}) Q_{mk}^{\alpha_2\alpha_3}(\bm{r}) Q_{km}^{\alpha_3\alpha_2}(\bm{r}) \rangle +3 \langle Q_{nn}^{\alpha_1\alpha_1}(\bm{r})\rangle \langle Q_{mk}^{\alpha_2\alpha_3}(\bm{r}) Q_{km}^{\alpha_3\alpha_2}(\bm{r}) \rangle
 \notag \\
 &
 +2 \langle Q_{nm}^{\alpha_1\alpha_2}(\bm{r}) Q_{mk}^{\alpha_2\alpha_3}(\bm{r}) Q_{kn}^{\alpha_3\alpha_1}(\bm{r}) \rangle +2 \langle Q_{nn}^{\alpha_1\alpha_1}(\bm{r}) \rangle \langle Q_{mm}^{\alpha_2\alpha_2}(\bm{r}) \rangle \langle  Q_{kk}^{\alpha_3\alpha_3}(\bm{r}) \rangle
 \end{align}
 and replica indices $\alpha_1$, $\alpha_2$ and $\alpha_3$ are all different. In the two-loop approximation we find
 \begin{gather}
 P_3^{\alpha_1\alpha_2\alpha_3}(i\varepsilon_{n_1},i\varepsilon_{n_3},i\varepsilon_{n_2}) = P^{\alpha_1\alpha_2\alpha_3}_3(i\varepsilon_{n_1},i\varepsilon_{n_2},i\varepsilon_{n_3}) = P^{\alpha_1\alpha_2\alpha_3}_3(i\varepsilon_{n_2},i\varepsilon_{n_1},i\varepsilon_{n_3})= - P^{\alpha_1\alpha_2\alpha_3}_3(i\varepsilon_{n_2},i\varepsilon_{n_4},i\varepsilon_{n_1})  \notag \\
 = - P^{\alpha_1\alpha_2\alpha_3}_3(i\varepsilon_{n_2},i\varepsilon_{n_1},i\varepsilon_{n_4})
 = - P^{\alpha_1\alpha_2\alpha_3}_3(i\varepsilon_{n_1},i\varepsilon_{n_2},i\varepsilon_{n_4}) = - \left [ \frac{4}{g} \int_p D_p(0) \right ]^2 = - \left [ \frac{8 \Omega_d h^\epsilon}{\epsilon g}\right ]^2+O(1)
 \notag \\
 P^{\alpha_1\alpha_2\alpha_3}_3(i\varepsilon_{n_1},i\varepsilon_{n_3},i\varepsilon_{n_5}) = P^{\alpha_1\alpha_2\alpha_3}_3(i\varepsilon_{n_2},i\varepsilon_{n_4},i\varepsilon_{n_6}) = 0 .
 \end{gather}
Hence, we obtain
\begin{equation}
K_3 = \rho_0^3 \frac{3 t^2 h^{2\epsilon}}{\epsilon^2} +O(1) .
\end{equation}
By using Eq. \eqref{eqmqe3}, we obtain Eq. \eqref{eqmqe}.

\subsection{C.\, The $4$-th irreducible moment of the LDOS}

The $4$-th irreducible moment $K_4$ can be obtained from the function
 \begin{align}
 K_4 &= \frac{\rho_0^4}{16} \Bigl (P^{\alpha_1\alpha_2\alpha_3\alpha_4}_4(i\varepsilon_{n_1},i\varepsilon_{n_3},i\varepsilon_{n_5},i\varepsilon_{n_7}) - P^{\alpha_1\alpha_2\alpha_3\alpha_4}_4(i\varepsilon_{n_1},i\varepsilon_{n_3},i\varepsilon_{n_5},i\varepsilon_{n_2}) 
 - P^{\alpha_1\alpha_2\alpha_3\alpha_4}_4(i\varepsilon_{n_1},i\varepsilon_{n_3},i\varepsilon_{n_2},i\varepsilon_{n_5})
  \notag \\
&
 - P^{\alpha_1\alpha_2\alpha_3\alpha_4}_4(i\varepsilon_{n_1},i\varepsilon_{n_2},i\varepsilon_{n_3},i\varepsilon_{n_5}) 
 - P^{\alpha_1\alpha_2\alpha_3\alpha_4}_4(i\varepsilon_{n_2},i\varepsilon_{n_1},i\varepsilon_{n_3},i\varepsilon_{n_5})  
 +P^{\alpha_1\alpha_2\alpha_3\alpha_4}_4(i\varepsilon_{n_1},i\varepsilon_{n_3},i\varepsilon_{n_2},i\varepsilon_{n_4}) 
   \notag \\
&
 +P^{\alpha_1\alpha_2\alpha_3\alpha_4}_4(i\varepsilon_{n_1},i\varepsilon_{n_2},i\varepsilon_{n_3},i\varepsilon_{n_4}) 
  +P^{\alpha_1\alpha_2\alpha_3\alpha_4}_4(i\varepsilon_{n_2},i\varepsilon_{n_1},i\varepsilon_{n_3},i\varepsilon_{n_4}) 
 +P^{\alpha_1\alpha_2\alpha_3\alpha_4}_4(i\varepsilon_{n_2},i\varepsilon_{n_4},i\varepsilon_{n_1},i\varepsilon_{n_3}) 
 \notag \\
&
+P^{\alpha_1\alpha_2\alpha_3\alpha_4}_4(i\varepsilon_{n_2},i\varepsilon_{n_1},i\varepsilon_{n_4},i\varepsilon_{n_3})
 +P^{\alpha_1\alpha_2\alpha_3\alpha_4}_4(i\varepsilon_{n_1},i\varepsilon_{n_2},i\varepsilon_{n_4},i\varepsilon_{n_3})
  - P^{\alpha_1\alpha_2\alpha_3\alpha_4}_4(i\varepsilon_{n_2},i\varepsilon_{n_4},i\varepsilon_{n_6},i\varepsilon_{n_1}) 
 \notag \\
&
 - P^{\alpha_1\alpha_2\alpha_3\alpha_4}_4(i\varepsilon_{n_2},i\varepsilon_{n_4},i\varepsilon_{n_1},i\varepsilon_{n_6})
 - P^{\alpha_1\alpha_2\alpha_3\alpha_4}_4(i\varepsilon_{n_2},i\varepsilon_{n_1},i\varepsilon_{n_4},i\
 \varepsilon_{n_6}) 
 - P^{\alpha_1\alpha_2\alpha_3\alpha_4}_4(i\varepsilon_{n_1},i\varepsilon_{n_2},i\varepsilon_{n_4},i\varepsilon_{n_6})
   \notag \\
&
 +P^{\alpha_1\alpha_2\alpha_3\alpha_4}_4(i\varepsilon_{n_2},i\varepsilon_{n_4},i\varepsilon_{n_6},i\varepsilon_{n_8})
   \Bigl )  \label{eqK4P4}
 \end{align}
  after analytic continuation to the real frequencies: $\varepsilon_{n_{1,3,5,7}} \to E+i0^+$ and $\varepsilon_{n_{2,4,6,8}} \to E-i0^+$. Here
 \begin{align}
 P_4^{\alpha_1\alpha_2\alpha_3\alpha_4} (i\varepsilon_{n}, i\varepsilon_{m}, i\varepsilon_{k} ,i\varepsilon_{l})
  & = \langle Q_{nn}^{\alpha_1\alpha_1}(\bm{r}) Q_{mm}^{\alpha_2\alpha_2}(\bm{r}) Q_{kk}^{\alpha_3\alpha_3}(\bm{r}) Q_{ll}^{\alpha_4\alpha_4}(\bm{r}) \rangle - 4 \langle Q_{nn}^{\alpha_1\alpha_1}(\bm{r}) \rangle \langle  Q_{mm}^{\alpha_2\alpha_2}(\bm{r}) Q_{kk}^{\alpha_3\alpha_3}(\bm{r})Q_{ll}^{\alpha_4\alpha_4}(\bm{r}) \rangle 
 \notag \\
&+6 \langle Q_{nn}^{\alpha_1\alpha_1}(\bm{r}) Q_{ll}^{\alpha_4\alpha_4}(\bm{r}) \rangle \langle  Q_{mm}^{\alpha_2\alpha_2}(\bm{r}) Q_{kk}^{\alpha_3\alpha_3}(\bm{r}) \rangle
  -6 \langle Q_{nn}^{\alpha_1\alpha_1}(\bm{r}) Q_{ll}^{\alpha_4\alpha_4}(\bm{r})Q_{mk}^{\alpha_1\alpha_2}(\bm{r}) Q_{km}^{\alpha_2\alpha_1}(\bm{r}) \rangle 
  \notag \\
&
  +12 \langle Q_{nn}^{\alpha_1\alpha_1}(\bm{r}) \rangle \langle Q_{ll}^{\alpha_4\alpha_4}(\bm{r}) Q_{mk}^{\alpha_2\alpha_3}(\bm{r}) Q_{km}^{\alpha_3\alpha_2}(\bm{r}) \rangle 
    -6 \langle Q_{nn}^{\alpha_1\alpha_1}(\bm{r}) Q_{ll}^{\alpha_4\alpha_4}(\bm{r}) \rangle \langle  Q_{mk}^{\alpha_2\alpha_3}(\bm{r}) Q_{km}^{\alpha_3\alpha_2}(\bm{r}) \rangle
 \notag \\
&
 +8 \langle Q_{nn}^{\alpha_1\alpha_1}(\bm{r}) Q_{lm}^{\alpha_4\alpha_2}(\bm{r}) Q_{mk}^{\alpha_2\alpha_3}(\bm{r}) Q_{kl}^{\alpha_3\alpha_4}(\bm{r}) \rangle
  8 \langle Q_{nn}^{\alpha_1\alpha_1}(\bm{r}) \rangle \langle Q_{lm}^{\alpha_4\alpha_2}(\bm{r}) Q_{mk}^{\alpha_2\alpha_3}(\bm{r}) Q_{kl}^{\alpha_3\alpha_4}(\bm{r}) \rangle
 \notag \\
&
 +3 \langle Q_{nm}^{\alpha_1\alpha_2}(\bm{r}) Q_{mk}^{\alpha_2\alpha_3}(\bm{r}) Q_{kl}^{\alpha_3\alpha_4}
(\bm{r}) Q_{ln}^{\alpha_4\alpha_1}(\bm{r}) \rangle 
 - 6 \langle Q_{nm}^{\alpha_1\alpha_2}(\bm{r}) Q_{mn}^{\alpha_2\alpha_1}(\bm{r}) Q_{kl}^{\alpha_3\alpha_4}
(\bm{r}) Q_{lk}^{\alpha_4\alpha_3}(\bm{r}) \rangle 
\notag \\
&
-3  \langle Q_{nn}^{\alpha_1\alpha_1}(\bm{r}) \rangle \langle Q_{mm}^{\alpha_2\alpha_2}(\bm{r}) \rangle \langle  Q_{kk}^{\alpha_3\alpha_3}(\bm{r}) \rangle
\langle  Q_{ll}^{\alpha_4\alpha_4}(\bm{r}) \rangle 
 \end{align}
 and replica indices $\alpha_1$, $\alpha_2$, $\alpha_3$, $\alpha_4$ are all different.
  In the two-loop approximation we find that all $P_4$ in Eq. \eqref{eqK4P4} are zero except the following ones
 \begin{gather}
  P_4^{\alpha_1\alpha_2\alpha_3\alpha_4}(i\varepsilon_{n_1},i\varepsilon_{n_2},i\varepsilon_{n_3},i\varepsilon_{n_4}) 
 =P_4^{\alpha_1\alpha_2\alpha_3\alpha_4}(i\varepsilon_{n_2},i\varepsilon_{n_1},i\varepsilon_{n_3},i\varepsilon_{n_4}) 
  =P_4^{\alpha_1\alpha_2\alpha_3\alpha_4}(i\varepsilon_{n_2},i\varepsilon_{n_4},i\varepsilon_{n_1},i\varepsilon_{n_3}) \notag \\ 
=P_4^{\alpha_1\alpha_2\alpha_3\alpha_4}(i\varepsilon_{n_2},i\varepsilon_{n_1},i\varepsilon_{n_4},i\varepsilon_{n_3}) 
=3
\left [ \frac{4}{g} \int_p D_p(0) \right ]^2 = 3 \left [ \frac{8 \Omega_d h^\epsilon}{\epsilon g}\right ]^2+O(1) .
 \end{gather}
Hence,
\begin{equation}
K_4 = \rho_0^4 \frac{3 t^2 h^{2\epsilon}}{\epsilon^2} +O(1) .
\end{equation}
By using Eq. \eqref{eqmqe3}, we obtain Eq. \eqref{eqmqe}.

\section{V.\, Scaling analysis}

Our results interpolate  between the case of non-interacting fermions ($\gamma_s=0$) and the case of fermions with Coulomb  interaction ($\gamma_s=-1$). In the case of short-range singlet interaction ($-1<\gamma_s<0$) the system flows towards the non-interacting limit $\gamma_s=0$ (see Eq. \eqref{betafunct}). In both cases, $\gamma_s=0$ and $\gamma_s=-1$, there are Anderson transitions in $d=2+\epsilon$ dimensions but they are of different universality classes, i.e. they have different sets of critical exponents. 

In the case of Coulomb interaction, $\gamma_s=-1$, renormalization of dimensionless conductance $g$ (in units $e^2/h$) is governed by the following $\beta$-function ($t=1/\pi g$):
\begin{equation}
-\frac{dt}{d\ln y}  = \beta(t,-1) = \epsilon t - 2 t^2 - 4 A t^3 + O(t^4) . \notag \\
\label{Scal_eq1}
\end{equation}
Here we add to Eq.~\eqref{betafunct} the two-loop contribution with the numerical constant $A$ given as [S9]
\begin{align}
A =\frac{1}{16}\left [\frac{139}{6}+\frac{(\pi^2-18)^2}{12}+\frac{19}{2}\zeta (3)+\Bigl ( 16 + \frac{\pi ^2}{3} \Bigr )\ln ^{2}2  - \Bigl (44-\frac{\pi ^{2}}{2}+7\zeta (3)\Bigr ) \ln 2+16\mathcal{G}-\frac{1}{3}\ln ^{4}2-8\lit\left(\frac{1}{2}\right)\right ] \approx 1.64 ,
\end{align}
where $\mathcal{G} \approx 0.915 $ denotes the Catalan constant, and $\lit(x) = \sum_{k=1}^\infty x^k/k^4$ denotes the polylogarithm. The zero of the $\beta$-function, $\beta(t_*,-1)=0$, determines the critical point
\begin{equation}
t_* = (\epsilon/2) (1-A\epsilon) +O(\epsilon^3) 
\end{equation}
at which the correlation (localization) length diverges
 \begin{equation}
\xi = l |t-t_*|^{-\nu},\qquad  \nu = -1/\beta^\prime(t_*,-1) = 1/\epsilon - A + O(\epsilon) .
 \end{equation}
Here $l$ is the microscopic scale at which $t$ is defined. Physically it corresponds to the elastic scattering mean free path. 

In the case of Coulomb interaction Eq.~\eqref{eqmqRG} becomes
\begin{equation}
-\frac{d \ln m_q}{d \ln y}  = \zeta_q(t) =  q(1-q)\frac{t}{2}  +q(1-q) \left ( 1 -\frac{\pi^2}{12}\right ) t^2 + O(t^3)  .
\label{Scal_eq1a}
\end{equation} 
Together with Eq.~\eqref{eqKmq}, it implies that at zero energy and temperature, $E=T=0$, the $q$-th moment of the LDOS obeys the following scaling behavior
\begin{equation}
\langle \rho^q \rangle \sim \langle \rho \rangle^q \left ({\xi}/{l}\right )^{-\Delta_q} \Upsilon_q(\xi/L) .
\label{Scal_eq3}
\end{equation}
Here $L$ stands for the system size, the multifractal critical exponent $\Delta_q$ is determined by the anomalous dimension of $m_q$ at the critical point: 
 \begin{equation}
\Delta_q = \zeta_q(t_*,-1) =
\frac{q(1-q)\epsilon}{4}\Bigl [1 + \left (1-A-\frac{\pi^2}{12}\right ) \epsilon \Bigr ]+O(\epsilon^3) ,
\end{equation}
and the scaling function $\Upsilon_q(x)$ has the following asymptotes
\begin{equation}
\Upsilon_q(x) = \begin{cases}
1, & \qquad x\ll 1 \\
x^{\Delta_q}, & \qquad x\gg 1 .
\end{cases} 
\label{Scal_eq10}
\end{equation}
 
As one can see from Eq. \eqref{Scal_eq3}, the scaling behavior of the $q$-th moment of the LDOS is determined also by the scaling behavior of the average DOS. At zero energy and temperature, $E=T=0$, one can write [S2,S10] 
\begin{equation}
\langle \rho \rangle \sim (\xi/l)^{-\theta} \mathcal{K}(\xi/L), 
\label{Scal_eq2}
\end{equation}
where the critical exponent $\theta=1+O(\epsilon)$ and the scaling function $\mathcal{K}$ behaves as follows
 \begin{equation}
\mathcal{K}(x) = \begin{cases}
1, & \qquad x\ll 1 , \\
x^{\theta}, & \qquad x\gg 1 .
\end{cases} 
\end{equation}
Combining Eqs \eqref{Scal_eq3} and \eqref{Scal_eq2}, we find
\begin{equation}
\langle \rho^q \rangle \sim  \left ({\xi}/{l}\right )^{-\theta q-\Delta_q} \tilde{\Upsilon}_q(\xi/L) ,
\label{Scal_eq4}
\end{equation}
where the scaling function $\tilde\Upsilon_q(x)$ has the following properties:
\begin{equation}
\tilde\Upsilon_q(x) = \begin{cases}
1, & \qquad x\ll 1 \\
x^{\Delta_q+\theta q}, & \qquad x\gg 1 .
\end{cases} 
\end{equation}
Since the multifractal exponent is negative, $\Delta_q<0$, the combination $\theta q+\Delta_q$ can be of arbitrary sign in general.

As usual, in the presence of interactions, finite energy or temperature induces the inelastic length $L_\phi$ related with the  dephasing time $\tau_\phi$: $L_\phi \sim \tau_\phi^{1/z}$, where $z$ is the dynamical exponent. In the case of Coulomb interaction the frequency/energy and temperature scaling are the same such that $1/\tau_\phi \sim \max\{|E|,T\}$. Therefore, the inelastic length  becomes $L_\phi \sim \min\{L_E, L_T\}$ with $L_E\sim |E|^{-1/z}$ and $L_T\sim T^{-1/z}$. We emphasize that the energy $E$ is counted from the chemical potential. The dynamical exponent $z$ is known up to the two-loop order [S3]: 
\begin{equation}
z=2+\frac{\epsilon}{2}+\left (2A-\frac{\pi^2}{6}-3\right )\frac{\epsilon^2}{4}+O(\epsilon^3) .
\end{equation}
Provided $L_\phi \ll L$, the inelastic length should be substituted for $L$ in Eqs~\eqref{Scal_eq3}, \eqref{Scal_eq2} and \eqref{Scal_eq4}. Therefore, our scaling results for the $q$-th moment can be summarized as follows:
\begin{equation}
\langle \rho^q(E,\bm{r}) \rangle \sim \langle \rho(E) \rangle^q \left (\mathcal{L}/{l}\right )^{-\Delta_q}  , \qquad \langle \rho(E) \rangle \sim \left (\mathcal{L}/{l}\right )^{-\theta} ,
\label{Scal_eq5}
\end{equation}
where $\mathcal{L} = \min\{L,\xi,L_\phi\}$. Note that the exponent $\theta$ is related with the exponent $\beta$ which determines the energy dependence of the average LDOS at the criticality, $\langle \rho(E)\rangle \sim |E|^{\beta}$, as $\beta= \theta/z$. We assume that $t \leqslant t_*$, i.e., the system is either exactly in the transition point or slightly on the metallic side. In the case of an insulator, $t > t_*$, a glassy behavior occurs. Investigation of LDOS fluctuations in this regime remains a prospect for future research.

The scaling behavior of the correlation function of the LDOS at the same energy but different spatial points can be 
written as
\begin{equation}
\langle \rho(E,\bm{r}) \rho(E,\bm{r}+\bm{R})\rangle \sim \langle \rho(E) \rangle^2 \left (R/\mathcal{L} \right )^{-\eta} .
\label{Scal_eq6}
\end{equation}
Here the exponent $\eta=-\Delta_2$ and it is assumed that $l\ll R \ll \mathcal{L}$. This result follows from two observations: i) at $R\sim l$ Eq. \eqref{Scal_eq6} should reproduce Eq.~\eqref{Scal_eq5} with $q=2$, ii) at $R\sim \mathcal{L}$ the LDOS at points $\bm{r}$ and $\bm{r}+\bm{R}$ is uncorrelated. Similarly (see, e.g. Ref. [\onlinecite{Evers08}]), one can find at $l\ll R \ll \mathcal{L}$ that  
\begin{equation}
\langle \rho^{q_1}(E,\bm{r}) \rho^{q_2}(E,\bm{r}+\bm{R})\rangle \sim \langle \rho(E) \rangle^{q_1+q_2} \left (R/\mathcal{L} \right )^{\Delta_{q_1+q_2}-\Delta_{q_1}-\Delta_{q_2}} \mathcal{L}^{-\Delta_{q_1}-\Delta_{q_2}} .
\end{equation}

The 2-point correlation function of the LDOS at the same spatial point but at different energies demonstrates the following scaling behavior:
\begin{gather}
\langle \rho(E,\bm{r})  \rho(E+\omega,\bm{r}) \rangle \sim \langle \rho(E) \rangle
\langle \rho(E+\omega) \rangle
 \left (L_\omega/{l}\right )^{\eta} \hat{\Upsilon}_2(L_\omega/L_E) ,\notag \\ 
  \langle \rho(E) \rangle \sim (L_E/l)^{-\theta}, \qquad 
\langle \rho(E+\omega) \rangle \sim  (\min\{L_E,L_\omega\}/l)^{-\theta} ,
\label{Scal_eq11}
\end{gather}
where the scaling function  $\hat{\Upsilon}_2(x)$ has the same asymptotes as the function  $\Upsilon_2(x)$    (see Eq. \eqref{Scal_eq10}). We assume here that the following condition holds $L_\omega, L_E \ll \min\{L, \xi, L_T\}$.
In the case $L_\omega \ll \tilde{\mathcal{L}} = \min\{L, \xi, L_T, L_E\}$, one expects the following scaling behavior:
\begin{equation}
\langle \rho(E,\bm{r})  \rho(E+\omega,\bm{r}) \rangle \sim (\tilde{\mathcal{L}}/l)^{-\theta}
 \left (L_\omega/{l}\right )^{-\theta+\eta} .
 \label{Scal_eq12}
\end{equation}

By using Eqs \eqref{Scal_eq6} and \eqref{Scal_eq11}, we find the following scaling behavior of the 2-point correlation function of the LDOS at different energies and different spatial points in the most interesting case $l\ll R\ll L_\omega\ll \tilde{\mathcal{L}}$:
\begin{gather}
\langle \rho(E,\bm{r})  \rho(E+\omega,\bm{r}+\bm{R}) \rangle \sim \langle \rho(E) \rangle
\langle \rho(E+\omega) \rangle
 \left (L_\omega/{R}\right )^{\eta}.
\label{Scal_eq14}
\end{gather}

Finally, it is instructive to compare the scaling results obtained above  for the case of Coulomb interaction with 
the multifractality at the Anderson transition in the system of fermions with short-range singlet interaction ($-1<\gamma_s\leqslant 0$). In fact, all the scaling results \eqref{Scal_eq3}, \eqref{Scal_eq2}, \eqref{Scal_eq4}, \eqref{Scal_eq5}, \eqref{Scal_eq6}, and \eqref{Scal_eq11} - \eqref{Scal_eq14} remain the same, but the critical exponents become different. As was mentioned above, the Anderson transition in the case of short-range singlet interaction is described by the $\beta$-function of the non-interacting ($\Gamma_s=0$) problem  [S11]
\begin{equation}
-\frac{dt}{d\ln y} = \beta(t,0) = \epsilon t - \frac{1}{2} t^3 - \frac{3}{8} t^5 + O(t^6)  .
\end{equation}
The critical point and correlation length exponent are given as
\begin{equation}
t_* = (2\epsilon)^{1/2}\left (1-\frac{3\epsilon}{4}\right )+O(\epsilon^{5/2}), \qquad \nu = \frac{1}{2\epsilon} - 
\frac{3}{4} +O(\epsilon) .
\end{equation}
The anomalous dimensions of $m_q$ were computed up to  the four-loop order [S12]:
\begin{equation}
\zeta_q(t,0) = \frac{q(1-q) t}{2}\left ( 1+ \frac{3 t^2}{8} + \frac{3 \zeta(3)}{16} q(q-1) t^3\right )+ O(t^5) ,
\end{equation}
where $\zeta(3)\approx 1.2$ stands for the Riemann zeta.
This leads to the following expression for the multifractal exponents in the non-interacting case: 
\begin{equation}
\Delta_q = q(1-q) \left (\frac{\epsilon}{2}\right )^{1/2} - \frac{3 \zeta(3)}{32} q^2(q-1)^2 \epsilon^2  +  O(\epsilon^{5/2}).
\end{equation}
The average LDOS is uncritical across the Anderson transition for non-interacting fermions such that the critical exponent $\theta=0$.  Finally, we remind that in the case of short-range singlet interaction length scales induced by finite temperature and frequency/energy are governed by different critical exponents [S1,S13,S14]: $L_E \sim |E|^{-1/z}$ and $L_T \sim T^{-1/z_T}$, with
\begin{equation}
z = d, \qquad z_T = 2 -2 (2\epsilon)^{1/2}+5\epsilon - 2 (2\epsilon)^{3/2} + O(\epsilon^2) .
\end{equation}

\begin{itemize}

\item[]
\item[] 
\item[] 

\item[[S1\!\!]] D. Belitz, T.R. Kirkpatrick, Rev. Mod. Phys. {\bf 66}, 261 (1994).

\item[[S2\!\!]]  A.M. Finkel'stein, vol. 14 of Soviet Scientific Reviews, ed. by I.M.\, Khalatnikov, Harwood Academic Publishers, London, (1990).

\item[[S3\!\!]] M.A. Baranov, A.M.M. Pruisken, and B. \v{S}kori\'{c}, Phys. Rev. B {\bf 60}, 16821 (1999).

\item[[S4\!\!]] B.L.\,Altshuler and A.G.\,Aronov, in
{\it Electron-Electron Interactions in Disordered Conductors}, ed.
A.J. Efros and M. Pollack, Elsevier Science Publishers,
North-Holland, 1985.

\item[[S5\!\!]] B.L. Altshuler, A.G. Aronov, and P.A. Lee, Phys. Rev. Lett. {\bf 44}, 1288 (1980).

\item[[S6\!\!]] A.M. Finkelstein, Sov. Phys. JETP {\bf 57}, 97 (1983).

\item[[S7\!\!]] C.\
Castellani, C.\ DiCastro, P.A. Lee and M.\ Ma, Phys.
Rev. B \textbf{34}, 527 (1984).

\item[[S8\!\!]] F.~Wegner, Z. Physik B {\bf 36}, 209 (1980).
 
\item[[S9\!\!]] M.A.~Baranov, I.S.~Burmistrov, and A.M.M.~Pruisken, Phys. Rev. B {\bf 66}, 075317 (2002).

\item[[S10\!\!]] D. Belitz, T.R. Kirkpatrick, Phys. Rev. B {\bf 48}, 14072 (1993).

\item[[S11\!\!]] S. Hikami, Nucl. Phys. B{\bf 215}, 555 (1983); W. Bernreuther and F.J. Wegner, Phys. Rev. Lett. {\bf 57}, 1383 (1986).

\item[[S12\!\!]] D. H\"of, F. Wegner, Nucl. Phys. B {\bf 275}, 561
  (1986); F. Wegner, Nucl. Phys. B {\bf 280}, 193
  (1987); Nucl. Phys. B {\bf 280}, 210 (1987).

\item[[S13\!\!]] F.~Evers and A.D.~Mirlin,
  Rev. Mod. Phys. {\bf 80},   1355 (2008).
  
\item[[S14\!\!]]  I.S. Burmistrov, S. Bera, F. Evers, I.V. Gornyi, and A.D. Mirlin, Ann. Phys. (N.Y.) {\bf 326}, 1457 (2011).

\end{itemize}
\end{widetext}


\begin{thebibliography}{99}
\bibitem{note2} The multifractality of non-interacting electrons may have a
dramatic impact on properties of problems with RG-relevant short-range
electron-electron interaction by determining the RG evolution of the system away
from the non-interacting fixed point. In particular, it was shown that the
multifractality leads to a strong enhancement of superconducting transition
temperature \cite{feigelman07,burmistrov12}  and  controls instability of
surface states of a topological superconductor with respect to interaction
\cite{foster12}.
\bibitem{note1} It is worth noting an interesting similarity between the
results (\ref{e3}), (\ref{e4}) and the behavior of LDOS at critical points
in non-interacting systems of unconventional symmetry classes (see
Ref. \cite{Evers08} and references therein). However, the physics in the two
cases is essentially different. In unconventional classes, the
LDOS is suppressed at a certain point of the spectrum of a single-particle
Hamiltonian where an additional symmetry arises. A representative example is
provided by models of Dirac fermions subjected to special types of disorder.
Contrary to this, in the problem we are considering the suppression of average
LDOS takes place because of Coulomb interaction and is bound to the chemical
potential. This suppression is a genuine many-body effect that has
common roots with formation of gap in Mott insulators and in Coulomb-blockade
regime of quantum dots, as well as of soft Coulomb gap \cite{ES75SE84} in disordered
insulators.
\end{thebibliography}
\end{document}